\documentclass{article}
\usepackage{graphicx}
\usepackage{epsfig}
\usepackage{amsfonts}
\usepackage{amssymb}
\usepackage{color}
\usepackage{epsf}
 \usepackage{amssymb}
\usepackage{enumitem}
\usepackage[english]{babel}
\usepackage[nottoc]{tocbibind}
\usepackage{caption}
\usepackage{subcaption}

\date{\today}

\newcommand{\insertplot}[5]{\begin{figure}
 \hfill\hbox to 0.05in{\vbox to #5in{\vfill
 \inputplot{#1}{#4}{#5}}\hfill}
 \hfill\vspace{-.1in}
 \caption{#2}\label{#3}
 \end{figure}}
 \newcommand{\inputplot}[3]{
 \special{ps: plotfile #1}
}
\newcounter{fig}

\newcommand{\ee}{\end{equation}}
\newcommand{\eea}{\end{eqnarray}}
\newcommand{\be}{\begin{equation}}
\newcommand{\bea}{\begin{eqnarray}}

\begin{document}

\title{\Large{\bf
Two boson stars in equilibrium
%Dipole Boson Stars:
%
%construction and physical properties
} }

\vspace{1.5truecm}
	
\author{
{\large }%$^{\ddagger}$
{\  P. Cunha}$^{1}$,
{\  C. Herdeiro}$^{1}$, 
{  E. Radu}$^{1}$
and
{ Ya. Shnir}$^{2,3}$
\\
\\
$^{1}${\small  Departamento de Matemática da Universidade de Aveiro}
\\
 {\small 
Center for Research and Development in Mathematics and Applications -- CIDMA}
\\
 {\small 
Campus de Santiago, 3810-183 Aveiro, Portugal 
}
\\  
$^{2}${\small
BLTP, JINR,
Joliot-Curie 6, Dubna 141980, Moscow Region, Russia}\\
$^{3}${\small
Institute of Physics, University of Oldenburg,
Oldenburg D-26111, Germany
}
}

\date{September 2022}

\maketitle

\begin{abstract}
We construct and explore the solution space of two non-spinning, mini-boson stars in equilibrium, in fully non-linear General Relativity (GR), minimally coupled to a free, massive, complex scalar field. The equilibrium is due to the balance between the (long range) gravitational attraction and the (short-range) scalar mediated repulsion, the latter enabled by a $\pi$ relative phase. Gravity is \textit{mandatory}; it is shown no similar solutions exist in flat spacetime, replacing gravity by non-linear scalar interactions. We study the variation of the proper distance between the stars with their mass (or oscillation frequency), showing it can be qualitatively captured by a simple analytic model that features the two competing interactions. Finally, we discuss some physical properties of the solutions, including their gravitational lensing.
\end{abstract}

\newpage

\tableofcontents

 %%%%%%%%%%%%%%%%%%%%%%%%%%%%%%%%%%%%%%%%%%%
\section{Introduction}
 %%%%%%%%%%%%%%%%%%%%%%%%%%%%%%%%%%%%%%%%%%%
The two-body equilibrium is an old and far reaching problem in General Relativity (GR). As a non-linear theory, there is no superposition principle. Thus, the putative existence of any two-body equilibrium solution must be anchored on a balance between different interactions. 

It was therefore surprising when Bach and Weyl, following Weyl's formalism for static, axi-symmetric solutions in GR~\cite{Weyl:1917gp}, found a two-body equilibrium solution in fully non-linear \textit{vacuum} GR. This fact intrigued Einstein, who in 1936 addressed the problem with Rosen~\cite{Einstein:1936fp}, realizing there is indeed an extra (non-vacuum) ingredient that allows the equilibrium: a conical singularity. The latter  can be either interpreted as a strut in between the two bodies, or two strings connecting each body to infinity. Whichever the chosen interpretation, the solution is geodesically incomplete, by virtue of the (naked) conical singularity. 

A non-singular (on and outside the event horizon) two-body (or in fact $N$-body) solution of GR was eventually found, independently, in 1947~\cite{Majumdar:1947eu} by Majumdar and in 1948 by 
Papapetrou~\cite{Papapetrou:1948jw} in \textit{electro-vacuum}. They found that, under an appropriate ansatz, the Einstein-Maxwell equations \textit{fully} linearise into a Laplace equation in flat Euclidean 3-space, which admits a multi-centre harmonic solution. Each centre (in the full geometry) was later interpreted by Hartle and Hawking~\cite{Hartle:1972ya} as the horizon of an extremal Reissner-Nordstr\"om black hole (BH).

The existence of the two-centre (or multi-centre) Majumdar-Papapetrou solution can be interpreted as due to a balance between the gravitational attraction and the electric repulsion. In classical mechanics, two point masses of magnitude $M$, each electrically charged with magnitude $Q$, at a distance $r$, interact via a potential energy $U=U_g+U_e$, where 
\begin{equation}
U_g=-G\frac{M^2}{r} \ , \qquad U_e=\frac{1}{4\pi \epsilon_0}\frac{Q^2}{r} \ .
\end{equation}
In geometrized units, for which Newton's constant $G$ and the vacuum electric permittivity $\epsilon_0$ obey, $G=1=4\pi \epsilon_0$, extremal objects obeying $M=Q$ can be in equilibrium at \textit{any} $r$, $i.e.$ ${\bf F}=-\nabla U=0$. This is precisely what occurs for the Majumdar-Papapetrou solution, wherein the extremal BHs can be placed at any location. Thus, this heuristic classical mechanics argument provides an intuition for the Majumdar-Papapetrou solution. This argument also suggests the equilibrium is not stable against small perturbations -- there is a flat potential -- which in fact is also the case in the full GR problem~\cite{Gibbons:1986cp,Ferrell:1987gf}. Small perturbations lead to an interesting scattering problem which can be dealt, for small velocities, in the moduli space approximation - see also~\cite{Manton:1988ba}.

\medskip

The simplest extension of vacuum GR, apart from electro-vacuum, is argueably scalar-vacuum. Allowing the scalar field to be massive and complex, but still free and minimally coupled to gravity, yields a novel feature: horizonless self-gravitating solitons exist, describing localized energy lumps, dubbed \textit{boson stars} (BSs). These solutions were first described in 1968 by Kaup~\cite{Kaup:1968zz}, in 1969 by Ruffini and Bonazzala \cite{Ruffini:1969qy} and (in a less known work) in 1968 by Feinblum and McKinley~\cite{Feinblum:1968nwc}. Single BSs rely on a dispersive scalar field -- due to an oscillating amplitude with a harmonic phase $\omega$, but with a time independent energy-momentum tensor -- being confined by gravity. The harmonic phase $\omega$ can neither be  too large, with an upper bound set by the field's mass $\mu$, nor too small, with a minimum frequency $\omega_{\rm min}$. For $\omega_{\rm min}<\omega<\mu$, self-gravitating, everywhere regular, asymptotically flat spherical solitons exist, for one or more values of the ADM mass (in units of the field's mass) $M\mu$, defining one or more branches - see $e.g.$~\cite{Herdeiro:2017fhv}.\footnote{Here we are referring to the \textit{fundamental} solutions, for which the scalar field has no nodes. Excited, spherical solutions with nodes also exist - see $e.g.$~\cite{Sanchis-Gual:2021phr}. Moreover, spinning, axially-symmetric solutions also exist, see $e.g.$~\cite{Herdeiro:2019mbz}, which can be seen as a different sort of excitation, with angular (rather than radial) nodes.} Some of these solutions are dynamically robust -- see ~\cite{Schunck:2003kk,Liebling:2012fv,Shnir:2022lba} for reviews -- and one may ask if one can set two such boson stars in equilibrium in this fully non-linear theory.

Non-linear BS solutions that are axi-symmetric, static, asymptotically flat and describe two symmetric lumps of scalar field energy were first reported in~\cite{Yoshida:1997nd} (see also \cite{Schupp:1995dy,Yoshida:1997jq} for a discussion in the Newtonian limit). More recently, more complex configurations of many scalar lumps in equilibrium were reported~\cite{Herdeiro:2020kvf}\footnote{See also~\cite{Herdeiro:2021mol,Gervalle:2022fze} for chains of non-spinning and spinning BSs.}, therein dubbed \textit{multipolar BSs}. Although these are non-linear configurations, and therefore not a simple superposition of two BSs, it is natural to interpret the two-centre \textit{dipolar} BSs (DBSs) as an equilibrium solutions of two (equal mass) BSs~\cite{Yoshida:1997nd}.

A key finding in the works described in the previous paragraph is that DBSs solutions exist only for a parity odd scalar field, with respect to the symmetry plane in between the two individual BSs. This is equivalent to saying the two BSs have a phase difference of $\pi$. Such a phase difference sustains a \textit{repulsive} scalar interaction, as confirmed (say) by performing head-on collisions of BSs, using numerical relativity techniques~\cite{Palenzuela:2006wp}. Since the scalar field has mass $\mu$, one may expect to capture the leading interaction between the two BSs by an interaction potential including both gravity and a meson Yukawa term $U=U_g+U_s$, with
\begin{equation}
U_g=-G\frac{M^2}{r} \ , \qquad U_s=g\frac{Q^2 }{r}e^{-\alpha \mu r} \ ,
\label{emodel}
\end{equation}
where $Q$ defines the scalar strength of each BS -- its \textit{Noether charge} -- and $g,\alpha$ are constants. This heuristic model suggests that: 1) For each pair of BSs with given $M$ (and $Q=Q(M)$) there is a specific equilibrium distance $L=L[M,Q(M)]$; 2) This equilibrium is stable, at least in the point particle approximation. Below we shall discuss how this simple model indeed captures the distance dependence of the fully non-linear DBSs and why the stability problem of DBSs is more complex than what this model suggests.  
\medskip

In this paper we shall construct the domain of solutions of DBSs and explore its physical properties, within the perspective that it is the simplest multi-BSs configuration. This serves both as a test ground for the properties of equilibrium multi-BSs and as a bridge towards the dynamical problem of collisions of two BSs. We shall further comment on both these perspectives in the final discussion. 

This paper is organized as follows. In Section~\ref{framework} we introduce the model, the ansatz and the explicit equations of motion to construct DBSs. In Section~\ref{sec_bc_num} we discuss the boundary conditions under which the equations of motion are solved and introduce some relevant physical quantities for the analysis of the solutions. In Section~\ref{sec_numerics} we present the numerical approach to solve the equations of motion. In Section~\ref{sec_res} we discuss the solution space and illustrate specific solutions. Moreover, we discuss how a simple analytic model based on~(\ref{emodel}) captures (to some extent) the behaviour of the fully non-linear solutions. We also provide an argument for the absence of two static $Q$-balls in equilibrium on Minkowski spacetime, regardless of the specific self-interactions.  In Section~\ref{sec-lensing} we describe the lensing properties of the DBSs and we conclude with a discussion and final remarks in Section~\ref{sec-final}.

%%%%%%%%%%%%%%%%%%%%%%%%%%%%%%%%%%%%%%%%%%%%%%%%%%%%%%%%%%%%%%%%%%%%%%%%%%%%%%%
\section{The  framework}
\label{framework}
%%%%%%%%%%%%%%%%%%%%%%%%%%%%%%%%%%%%%%%%%%%%%%%%%%%%%%%%%%%%%%%%%%%%%%%%%%%%%%%%

%%%%%%%%%%%%%%%%%%%%%%%%%%%%%%%%%%%%%%%%%%%%%%%%%%%%%%%%%%%%%%%%%%%%%%%%%%%%%%%
\subsection{The model and the ansatz}
\label{sec_eq_motion}
%%%%%%%%%%%%%%%%%%%%%%%%%%%%%%%%%%%%%%%%%%%%%%%%%%%%%%%%%%%%%%%%%%%%%%%%%%%%%%%%

%%%%%%%%%%%%%%%%%%%%%%%%%%%%%%%%%%%%%%%%%%%%%%%%%%%%%%%%%%%%%%%%%%%%%%%%%%%%%%%
%\subsubsection{Action and field equations}
%\label{sec_eq_motion_action}
%%%%%%%%%%%%%%%%%%%%%%%%%%%%%%%%%%%%%%%%%%%%%%%%%%%%%%%%%%%%%%%%%%%%%%%%%%%%%%%%
 
We consider 
the action for Einstein's gravity minimally coupled to a complex massive scalar field  $\Phi $  
\begin{equation}
\label{action}
 S= \frac{1 }{16\pi G}\int  d^4x\sqrt{-g}\left[R  
  -\frac{1}{2} g^{\alpha\beta}\left( \Phi_{, \, \alpha}^* \Phi_{, \, \beta} + \Phi _
{, \, \beta}^* \Phi _{, \, \alpha} \right) - \mu^2 \Phi^*\Phi 
 \right] \ ,
\end{equation}
where $G$ is Newton's constant, $g_{\alpha\beta}$ the spacetime metric, with determinant $g$ and Ricci scalar $R$, ``*" denotes complex conjugation and $\mu$ is the scalar field  mass.

The resulting field equations are:
\begin{eqnarray}
\label{E-eq}
  &&
	E_{\alpha\beta}\equiv R_{\alpha\beta}-\frac{1}{2}g_{\alpha\beta}R-8 \pi G~T_{\alpha\beta}=0 \ , ~~
	\\
	&&
	\label{KG-eq}
\Box \Phi =\mu^2\Phi ,  
\end{eqnarray}
where 
\begin{equation}
T_{\alpha\beta} =  
2 \Phi_{ , (\alpha}^{ *}\Phi_{,\beta)} 
%+\Phi_{,b}^*\Phi_{,a} 
-g_{\alpha\beta} 
[  
 \Phi_{,\nu}^{*}\Phi^{,\nu}
 %+\Phi_{,d}^*\Phi_{,c} )
+\mu^2 \Phi^{*}\Phi] \ ,
\end{equation}
is the
 energy-momentum tensor  of the scalar field.

The action (\ref{action}) is invariant under the global $U(1)$ transformation 
$\Phi\rightarrow e^{i\alpha}\Phi$, where $\alpha$ is constant. 
This implies the existence  of a conserved current, $j^\nu=-i (\Phi^* \partial^\nu \Phi-\Phi \partial^\nu \Phi^*)$, with  $j^\nu_{\ ;\nu}=0$.
 It follows that integrating the timelike component of this 4-current in a spacelike slice $\Sigma$ 
yields a conserved quantity -- the \textit{Noether charge}:
\begin{eqnarray}
\label{Q}
Q=\int_{\Sigma}~j^t \ .
\end{eqnarray}
At a microscopic level, this Noether charge counts the number of scalar particles.

%%%%%%%%%%%%%%%%%%%%%%%%%%%%%%%%%%%%%%%%%%%
%\subsubsection{The Ansatz}
%%%%%%%%%%%%%%%%%%%%%%%%%%%%%%%%%%%%%%%%%%%

The line-element for the solutions to be constructed in this work  possesses  
two commuting Killing vector fields,
$\xi$ and $\eta$, with
\begin{equation}
\xi=\partial_t \ , \ \ \ \eta=\partial_{\varphi} \ ,
\end{equation}
in a system of adapted coordinates.
The \textit{analytical} study of GR solutions with these symmetries is usually considered
within a metric ansatz of the form
\begin{eqnarray}
\label{metric-cylindrical}
ds^2= -e^{-2U(\rho,z)} dt^2+e^{2U(\rho,z)}
 \Big[
e^{ 2k(\rho,z)} (d \rho^2+  dz^2) 
+P(\rho,z)^2 d\varphi^2
\Big]
\ ,
\end{eqnarray}
where $(\rho,z)$ correspond asymptotically to the usual cylindrical coordinates.
The corresponding expression for the scalar field is
\begin{eqnarray}
\label{scalar_ansatz}
\Phi=\phi(\rho,z)e^{-i \omega t}~,
\end{eqnarray} 
 where the real function
$\phi$ is the field amplitude
and
 $\omega$ 
is the scalar field frequency,
which we take to be positive, without loss of generality. 
The Einstein--Klein-Gordon (EKG) equations
take the following compact form\footnote{There are two further constraint equations
which, however, we do not display here.}:
\begin{eqnarray}
&&
\nonumber
 \nabla^2 k+(\nabla U)^2+8\pi G
\Big[
(\nabla \phi)^2+e^{2k+2U}(\mu^2-2e^{2U} \omega^2)\phi^2
\Big] =0 \ ,
\\
\label{eqE-culindrical}
&&
\nabla^2 U+\frac{1}{P}  (\nabla U)\cdot (\nabla P) 
-8\pi G e^{2k+2U} (\mu^2-2e^{2U} \omega^2)\phi^2=0 \ ,
\\
\nonumber
&&
\nabla^2 P +16\pi G e^{2k+2U} (\mu^2-2e^{2U} \omega^2)\phi^2=0 \ ,
\\
\nonumber
&&
 \nabla^2 \phi+\frac{1}{P}  (\nabla P)\cdot (\nabla \phi) -
e^{2k+2U} (\mu^2 - e^{2U} \omega^2)\phi=0 \ ,
\end{eqnarray}
where
we define
$$
 \nabla^2  A = \frac{\partial^2 A}{\partial \rho^2}+\frac{\partial^2 A}{\partial z^2} \ ,  \qquad 
 (\nabla A)\cdot (\nabla B)   = \frac{\partial  A}{\partial \rho }\frac{\partial  B}{\partial \rho }
+\frac{\partial  A}{\partial z}\frac{\partial  B}{\partial z}~.
$$
In the  (electro-)vacuum case, it is always possible 
to set $P\equiv \rho$,
such that only two independent metric functions appear in the equations,
and $(\rho,z)$ become  the canonical Weyl coordinates,
the system being integrable  \cite{Stephani:2003tm}.

However, setting $P\equiv \rho$ is not possible for the case of interest here and
no exact solutions appear to exist for nonzero $(\omega,\mu)$,
in which case the problem is solved numerically.
Then, it is convenient to use
`quasi-isotropic' spherical  coordinates $(r,\theta)$ instead of $(\rho,z)$,
with the usual transformation 
\begin{eqnarray} 
\label{ct}
 \rho=r\sin \theta,~~z=r\cos \theta~,
\end{eqnarray}
and  
 the usual coordinate ranges, $0\leq r<\infty$, $0\leq \theta \leq \pi$.
Also, in order 
 to make contact with our previous work 
 \cite{Herdeiro:2015gia},
we redefine 
$U=-F_0$,
$k=F_1+F_0$
and
$P=e^{F_2+F_0}$  in (\ref{metric-cylindrical}).
The line-element and scalar field become
\begin{eqnarray}
\label{ansatz}
ds^2=-e^{2F_0(r,\theta)} dt^2+e^{2F_1(r,\theta)}(dr^2+r^2 d\theta^2)
+e^{2F_2(r,\theta)}r^2 \sin^2\theta  d\varphi^2 \ , \qquad 
\Phi=\phi(r,\theta)e^{-i \omega t}~,
\end{eqnarray}
which  is the Ansatz used in this work in the numerical treatment of 
the problem. 
The Minkowski spacetime background is approached for $r\to \infty$, with $F_0=F_1=F_2=0$.
One also remarks that
the symmetry axis of the spacetime is given by $\eta=0$,
which corresponds to the $z$-axis, with $\theta=0,\pi$.

%%%%%%%%%%%%%%%%%%%%%%%%%%%%%%%%%%%%%%%%%%%%%%%%%%%%%%%%%%%%%%%%%%%%%%%%%%%%%%%
\subsection{The explicit equations of motion}
\label{sec_eq_motion_eom}
%%%%%%%%%%%%%%%%%%%%%%%%%%%%%%%%%%%%%%%%%%%%%%%%%%%%%%%%%%%%%%%%%%%%%%%%%%%%%%%%

Given the Ansatz (\ref{ansatz}), the
explicit form of the Klein-Gordon (KG) equation (\ref{KG-eq}) reads
\begin{eqnarray}
\label{eq-phi}
\phi_{,rr} +\frac{\phi_{,\theta \theta} }{r^2}  
 +\Big(F_{0,r}+F_{2,r}+\frac{2}{r}\Big)\phi_{,r}
+\frac{1}{r^2}  (F_{0,\theta}+F_{2,\theta}+\cot \theta) \phi_{, \theta}
+
e^{-2F_0+2F_1}
\left(
\omega^2
-\mu^2 e^{2F_1}
\right)
\phi=0\ . 
\end{eqnarray}
The metric functions $F_i$ ($i=1,2,3$) satisfy the following second order
partial differential equations (PDEs):  
\begin{eqnarray}
\label{eq-F0}
&&
F_{0,rr} +\frac{F_{0,\theta \theta} }{r^2}  
 +\Big(F_{0,r}+F_{2,r}+\frac{2}{r}\Big)F_{0,r}
+\frac{1}{r^2}  (F_{0,\theta}+F_{2,\theta}+\cot \theta) F_{0, \theta}
\\
\nonumber
&&
{~~~~~~~~~~~~~~~~~~~~~~~~~~~~~~~~~~~~~~~~~~~~~~~~~~~~~~~~~~~~~~~~~~~~~~~}
+
8\pi G
e^{2F_1}
\left(
\mu^2
-2 e^{-2F_0} \omega^2
\right)
\phi^2=0\ ,~{~~~~~}
\\
%%%%%%%%%%%%%%%%%%%%%%%%
\label{eq-F1}
&&
F_{1,rr} +\frac{F_{1,\theta \theta} }{r^2}  
-\Big(F_{2,r}+\frac{1}{r}\Big)F_{0,r}
+\frac{F_{1,r}}{r}
-\frac{1}{r^2}  (F_{2,\theta} +\cot \theta) F_{0, \theta}
\\
\nonumber
&&
{~~~~~~~~~~~~~~~~~~~~~~~~~~~~~~~~~~~~~~~~~~~~~~~~~~~~~~~~~~~~~~~~~~~~~~~}
+
8\pi G
e^{2F_1}
\left(
\phi_{,r}^2+\frac{\phi_{, \theta}^2 }{r^2}
+ e^{-2F_0+2F_1} \omega^2 \phi^2
\right)
=0 \ ,~{~~~~~}
\\
\label{eq-F2}
&&
F_{2,rr} +\frac{F_{2,\theta \theta} }{r^2}  +\frac{F_{0,r}}{r}
 +\Big(F_{0,r}+F_{2,r}+\frac{3}{r}\Big)F_{2,r}
+\frac{1}{r^2} 
               \Big[ 
 (F_{0,\theta} + F_{2,\theta} )  F_{2,\theta} 
+ (F_{0,\theta} + 2F_{2,\theta} ) \cot \theta
                 \Big] 
								\\
\nonumber
&&
{~~~~~~~~~~~~~~~~~~~~~~~~~~~~~~~~~~~~~~~~~~~~~~~~~~~~~~~~~~~~~~~~~~~~~~~}
+
8\pi G
e^{2F_1}  \mu^2 \phi^2=0\ .~{~~~~~}
\end{eqnarray}
In addition, there are also two constraint equations:
\begin{eqnarray}
\nonumber
&&
 F_{0,rr} +F_{2,rr} +(F_{0,r} -2F_{1,r} )F_{0,r} 
-(2F_{1,r} -F_{2,r} )F_{2,r} 
-\frac{1}{r}(F_{0,r}  +2F_{1,r} -F_{2,r} )
-\frac{1}{r^2 }
\Big[
 F_{0,\theta \theta} +F_{2,\theta \theta} 
\\
\label{Eq11}
&&
{~~~}
+( F_{0,\theta} -2F_{0,\theta}) F_{0,\theta}
+( F_{2,\theta} -2F_{1,\theta}) F_{2,\theta}
-2 \cot \theta ( F_{1,\theta} - F_{2,\theta}) 
\Big]
+16 \pi G \Big(  \phi_{0,r}^2 - \frac{ \phi_{,\theta}^2}{r^2 }\Big)
=0 \ , \ \ \ \ 
\\
\nonumber
&&
F_{0,r \theta} +F_{2,r \theta} +F_{0,r  } F_{0, \theta} +F_{2,r  } F_{2, \theta} 
- (F_{1,r  } F_{2, \theta} +F_{2,r  } F_{1, \theta} )
- (F_{1,r  } F_{0, \theta} +F_{0,r  } F_{1, \theta} )
\\
\label{Eq12}
&&
{~~~~~~~~}
-\frac{1}{r}
(  F_{0, \theta}+ F_{1, \theta}) 
-\cot \theta (  F_{1, r}- F_{2, r}) 
+16 \pi G \phi_{, r}\phi_{, \theta} =0 \ ,
\end{eqnarray}
which are not solved directly, being
 used to check the numerical accuracy of the results

The above equations  (\ref{eq-F0})-(\ref{Eq12})
are derived as follows.
The only nonzero components of the Einstein tensor are 
$E_t^t, E_r^r,E_\theta^\theta,E_\varphi^\varphi, E_r^\theta$. 
These five equations are divided into two groups: 
three of these equations are solved together with the KG equation (\ref{eq-phi}),
 yielding a coupled system of four PDEs on the four unknown functions.
 The remaining two Einstein equations are treated as constraints. 
More precisely, one takes  the following combinations of the Einstein equations:
 \begin{eqnarray}
 \nonumber
&&
E_r^r+E_\theta^\theta+E_\varphi^\varphi-E_t^t =0 \,~~~~ \Longrightarrow ~~~{\rm equation}~(\ref{eq-F0}),
\\
&&
\label{EKG-eqs}
E_r^r+E_\theta^\theta-E_\varphi^\varphi-E_t^t=0 \,~~~~ \Longrightarrow ~~~{\rm equation}~(\ref{eq-F1}),
\\
\nonumber
&&
E_r^r+E_\theta^\theta-E_\varphi^\varphi+E_t^t =0 \,~~~~ \Longrightarrow ~~~{\rm equation}~(\ref{eq-F2}),
\\
\nonumber
&&
E_r^r-E_\theta^\theta =0 \,~~~~ \Longrightarrow ~~~{\rm equation}~(\ref{Eq11}),
\\
\nonumber
&&
E_r^\theta =0 \,~~~~ \Longrightarrow ~~~{\rm equation}~(\ref{Eq12}).
\end{eqnarray}
%
%

%%%%%%%%%%%%%%%%%%%%%%%%%%%%%%%%%%%%%%%%%%%%%%%%%%%%%%%%%%%%%%%%%%%%%%%%%%%%%%%
\section{Boundary conditions and physical quantities of interest }
\label{sec_bc_num}
%%%%%%%%%%%%%%%%%%%%%%%%%%%%%%%%%%%%%%%%%%%%%%%%%%%%%%%%%%%%%%%%%%%%%%%%%%%%%%%%

In order to perform the numerical integration of the system of equations described 
in Section~\ref{framework}, appropriate boundary conditions must be imposed,
which implement
the conditions of  asymptotic flatness and regularity at $r=0$ and at the symmetry axis. These are now discussed in detail.

%%%%%%%%%%%%%%%%%%%%%%%%%%%
\subsection{Boundary conditions and asymptotic expansion }
\label{sec_abc}
%%%%%%%%%%%%%%%%%%%%%%%%%%%

The solutions
reported in this work
 are constructed subject to the following boundary conditions:
\begin{eqnarray}
&&
\nonumber
 \partial_r F_0|_{r=0} = \partial_r F_1|_{r=0} = \partial_r F_2|_{r=0}= \phi |_{r=0}=0 \ , 
\\
&&
\label{bc}
 F_0|_{r=\infty} =F_1|_{r=\infty} =F_2|_{r=\infty} =\phi|_{r=\infty} =0 \ ,
\\
&&
\nonumber
 \partial_\theta  F_0|_{\theta =0,\pi } = \partial_\theta   F_1|_{\theta =0,\pi } =  \partial_\theta   F_2|_{\theta =0,\pi } =
 \partial_\theta   \phi|_{\theta =0,\pi } =0 \ .
\end{eqnarray}
Moreover, the absence of conical singularities implies also that 
$F_1=F_2$
on the symmetry axis ($\theta=0,\pi$) (as implied by a constraint equation),
a condition which was not imposed, 
being satisfied as a result of implementing a consistent numerical scheme.

\medskip
 
An asymptotic expression of the solutions compatible 
with the above boundary conditions
can be found.
Starting with the small-$r$ exprssions,
the first terms in a power series expansion of $(F_i,\phi)$ are
\begin{eqnarray}
&&
\label{small-r}
\nonumber
F_0=f_{00}+ s_0(1-3\cos^2\theta )r^2+\mathcal{O}(r^4)\ , \\
&&
F_1=f_{10}+\big[s_1 +(s_0-2 s_1 -4\pi G d_1^2)~\cos^2\theta \big]r^2+\mathcal{O}(r^4) \ ,
\nonumber
\\
&&
\nonumber
F_2=f_{10}+ \frac{1}{3}
\Big[] 
s_0(5\cos^2\theta-2)+s_1(1-4\cos^2\theta)-4\pi G d_1^2 (1+2\cos2\theta)
\Big] r^2+\mathcal{O}(r^4) \ , \\
&&
\phi=d_1 r \cos \theta  +\mathcal{O}(r^3) \ ,
\end{eqnarray}
where $f_{00},f_{10},s_0,s_1,d_1$ are constants fixed by numerics.
A relatively simple expression can also be written for 
large $r$.
The asymptotic behaviour of the scalar field is of the form
\begin{equation}
\label{scalar_asy}
\phi(r,\theta)= f(\theta)  \frac{e^{-\sqrt{\mu^2-w^2}r}}{r}+\dots \ , \qquad 
 {\rm with}~~f(\theta)=\sum_{k=0}^{\infty}c_k P_{ 2k+1}(\cos \theta) \ ,
\end{equation} 
 where 
$P_{ 2k+1} (\cos \theta)$ are the   Legendre polynomials
($e.g.$ 
$P_1=\cos\theta$,
$P_3=\cos\theta(5\cos(2\theta)-1)/4$, $etc.$),
while $c_k$ are constants fixed by the numerics. 
Neglecting the scalar field
 contribution -- since it decays faster than any
power of $r$ --
the leading order terms in a power series expansion of the metric functions are: 
\begin{eqnarray}
\nonumber
&&
F_0= \frac{M}{r}+\frac{1}{12} \Big[1+p_1+p_2 (1+3\cos 2\theta) \Big]\Big(\frac{M}{r}\Big)^3+\dots \ ,
\\
\label{exp-inf}
&&
F_1=-\frac{M}{r}-\frac{1}{4} \Big[1+p_1+p_2 \cos 2\theta \Big]\Big(\frac{M}{r}\Big)^2
-\frac{1}{12}\Big[1+p_1+p_2(1+3\cos 2\theta)
\Big]\Big(\frac{M}{r}\Big)^3+\dots \ ,
 \\
 \nonumber
&&
 F_2=-\frac{M}{r}-\frac{1}{4} (1+p_1)\Big(\frac{M}{r}\Big)^2
 -\frac{1}{12}\Big[1+p_1+p_2(1+3\cos 2\theta)\Big]\Big(\frac{M}{r}\Big)^3+\dots \ ,
\end{eqnarray}
with $M$ the ADM mass of solutions and 
$p_1,p_2$
arbitrary constants.

The approximate expression of the solutions on the symmetry axis is more complicated, the first terms
in a small-$\theta$ expansion being (a similar expansion holds for $\theta\to \pi$):
\begin{eqnarray}
&&
F_0(r,\theta)=f_{00}(r)+f_{02}(r) \theta^2+\mathcal{O}(\theta)^4\ , \qquad 
F_1(r,\theta)=f_{10}(r)+f_{12}(r) \theta^2+\mathcal{O}(\theta)^4 \ ,~~
\\
&&
F_2(r,\theta)=f_{10}(r)+f_{22}(r) \theta^2+\mathcal{O}(\theta)^4\ , \qquad 
\phi(r,\theta)=\phi_0(r)+\phi_{2}(r) \theta^2+\mathcal{O}(\theta)^4 \ ,
\end{eqnarray}
where the essential data is encoded in the functions 
$f_{00}(r)$,
$f_{10}(r)$
and
$\phi_{0}(r)$
which result from the  numerics.
One finds $e.g.$
(where a prime denotes the derivative $w.r.t.$ 
the $r$-coordinate):
\begin{eqnarray}
&&
\nonumber
f_{02}=-\frac{1}{4}r 
\Big\{
r f_{00}''+f_{00}' \big[2+r(f_{00}'+f_{10}'\big])
\Big\}
-2\pi G e^{2(f_{10}-f_{00})} r^2 (e^{2f_{00}} \mu^2-2\omega^2)\phi_0^2 \ ,
\\
&&
f_{12}= -\frac{1}{2}r 
\Big\{
f_{10}'+
\frac{r}{2}
\Big[
f_{00}''+2f_{10}''+f_{00}'(f_{00}'-f_{10}')
+8\pi G (2 \phi_0'^2+e^{2f_{10}}\mu^2 \phi_0^2)
\Big]
\Big\}~,
\\
&&
\nonumber
f_{22}= -\frac{1}{6}r 
\Big\{
3f_{10}'-\frac{1}{2}r \Big[f_{00}'' -2f_{10}'' +(f_{00}'-2f_{10}')(f_{00}'+f_{10}')\Big]
+4 \pi G  e^{2(f_{10}-f_{00})} r (e^{2f_{00}} \mu^2+2\omega^2)\phi_0^2   
\Big\}~,
\\
&&
\nonumber
\phi_2=-\frac{1}{4}r
\Big\{
r \phi_0''+\Big[2+r(f_{00}'+f_{10}')\Big]\phi_0'
-e^{2(f_{10}-f_{00})} r (e^{2f_{00}} \mu^2-\omega^2)\phi_0   
\Big\} \ .
\end{eqnarray}
 
Also, for all solutions discussed in this work,
the scalar field changes sign $w.r.t.$
a reflection on the equatorial plane, $\theta=\pi/2$,
while the metric tensor remains invariant.
That is, the solutions posses the symmetry
$F_i(r,\theta)=F_i(r,\pi-\theta)$,
$\phi(r,\theta)=-\phi(r,\pi-\theta)$.
As a result, it is enough to consider 
the range $0\leq \theta \leq \pi/2$
for the angular variable; then the functions
$F_i$ and $\phi$
satisfy the following boundary conditions on the equatorial plane:
\begin{equation}
\partial_\theta F_i\big|_{\theta=\pi/2} = \phi\big|_{\theta=\pi/2} = 0 \ .
\end{equation}

%%%%%%%%%%%%%%%%%%%%%%%%%%%%%%%%%%%%%%%%%%%%%%%%%%%%%%%%%%%%
\subsection{Quantities of interest }
\label{sec_q}
%%%%%%%%%%%%%%%%%%%%%%%%%%%%%%%%%%%%%%%%%%%%%%%%%%%%%%%%%%%%
 
Our solutions are static,  globally regular and
 without an event horizon or conical singularities. 
Also, they approach asymptotically the Minkowski spacetime background. 
Then, their ADM mass $M$  
can be obtained from the respective Komar expression \cite{Wald:1984rg},
\begin{equation}
\label{komar}
{M} = \frac{1}{{4\pi G}} \int_{\Sigma}
 R_{\alpha \beta}n^\alpha\xi^\beta dV \ ,~
\end{equation}
where $\Sigma$ denotes an asymptotically flat spacelike hypersurface,
$n^\alpha$ is normal to $\Sigma$ (with $n_\alpha n^\alpha = -1$)
and
$dV$ is the natural volume element on $\Sigma$.
After replacing in (\ref{komar}) the Ricci tensor by the
energy-momentum tensor -- via the Einstein equations (\ref{E-eq}) --, one finds  
\begin{eqnarray}
 \label{komarM2}
M
= \, 2 \int_{\Sigma} \left(  T_{\alpha \beta} 
-\frac{1}{2} \, g_{\alpha\beta} \, T_{\nu}^{\nu}
 \right) n^{\alpha}\xi^{\beta} dV
=
 4\pi \int_{0}^\infty dr \int_0^\pi d\theta~r^2\sin \theta ~e^{F_0+2F_1+F_2}
 \left(
 \mu^2-2 e^{-2F_2}
\omega^2
 \right)\phi^2 \  .\
\end{eqnarray}
Alternatively, 
the ADM mass can be read off from 
the asymptotic sub-leading behaviour of the metric function $g_{tt}$
\begin{eqnarray}
\label{asym}
g_{tt} =-e^{2F_0}
=-1+\frac{2GM}{r}+  \dots \ .   
\end{eqnarray}   
In addition to mass, there is also 
a conserved  Noether charge, which is computed from (\ref{Q}):
\begin{eqnarray}
\label{Q-int}
Q=2\pi \int_{0}^\infty dr \int_0^\pi d\theta  
~r^2\sin \theta ~e^{F_0+2F_1+F_2} \omega\phi^2 \ .
\end{eqnarray}
These two physical quantities are  connected via the 'first law'
\begin{eqnarray}
\label{first-law}
dM=\omega  dQ  \ .
\end{eqnarray}

To see how `compact' the dipole BSs are,
one defines the {\it compactness}
\cite{Amaro-Seoane:2010pks}
\begin{equation}
{\rm Compactness}=   \frac{2M_{99}}{R_{99}} \ ,
\label{compactness}
\end{equation}
where 
 $R_{99}$ is the circumferential radius containing 99\% of the DBS mass, $M_{99}$,
with $R=e^{F_2}r$ for
 the Ansatz (\ref{ansatz}). 
The separation between
the two
individual BSs, $i.e.$ between the two individual components of the DBS, is given by the proper distance between them
\begin{eqnarray}
\label{Lz}
L_z= \int_{-z_0}^{0} dr e^{F_1(r,\pi)}+\int_{0}^{z_0} dr e^{F_1(r,0)}
=2\int_{0}^{z_0} dr e^{F_1(r,0)}~,
\end{eqnarray}
with $\pm z_0$ the position on the $z$-axis of a single star. As we shall see below, $z_0$ is given by the maximum of the corresponding energy density, which is reasonable, since individual spherical, fundamental BSs attain their maximum energy density at their centre.

%%%%%%%%%%%%%%%%%%%%%%%%%%%%%%%%%%%%%%%%%%%%%%%
\section{The numerical approach}
\label{sec_numerics}
%%%%%%%%%%%%%%%%%%%%%%%%%%%%%%%%%%%%%%%%%%%%%%%

%%%%%%%%%%%%%%%%%%%%%%%%%%%%%%%%%%%%%%%%%%%%%%%
%\subsection{The scaling}
%\label{sec_scaling}
%%%%%%%%%%%%%%%%%%%%%%%%%%%%%%%%%%%%%%%%%%%%%%%

Apart from Newton's constant,
the model  (\ref{action}) 
possess two input parameters, corresponding 
to the field mass and frequency.
However, the dependence on both $G$ and $\mu$
 disappears from the equations 
when
using
natural units set by $\mu$ and $G$,
$i.e.$
when taking the 
following
field
redefinition together with a scaling of both $r$ and $\omega$:
\begin{eqnarray}
\phi \to \frac{\phi}{\sqrt{4\pi G}} \ , \qquad 
r\to \frac{r}{\mu} \ , \qquad 
\omega \to \frac{\omega}{\mu} \ .
 \end{eqnarray} 
As such, we are left with a single input parameter, which is the (scaled)
field frequency $\omega$.
 Also, the global charges and all other quantities of interest are 
 expressed in units set by $\mu$ and $G$; however,
 in order to simplify the output,
  we set $G=1$ in what follows.
 
%%%%%%%%%%%%%%%%%%%%%%%%%%%%%%%%%%%%%%%%%%%%%%%
%\subsection{Solving the equations}
%\label{sec_solving}
%%%%%%%%%%%%%%%%%%%%%%%%%%%%%%%%%%%%%%%%%%%%%%%
In our approach, the Einstein--Klein-Gordon
 equations reduce to a set of four
coupled non-linear elliptic PDEs for the functions 
$(F_0, F_1, F_2; \phi)$,
which are displayed in Section~\ref{sec_eq_motion_eom}. 
These equations 
 have been solved numerically subject to the boundary conditions 
 introduced above.
Apart from these, there are two more constraint Einstein equations
(also presented in Section~\ref{sec_eq_motion_eom})
which are not solved in our numerics.
Following an argument originally proposed in \cite{Wiseman:2002zc},
one can, however, show that
the identities $\nabla_\nu E^{\nu r} =0$ and $\nabla_\nu E^{\nu \theta}=0$,
imply the Cauchy-Riemann relations
%\begin{eqnarray}
$
\partial_{\bar r} {\cal P}_2  +
\partial_\theta {\cal P}_1
= 0 ,
$
$
 \partial_{\bar r} {\cal P}_1
-\partial_{\theta} {\cal P}_2
~= 0 .
$
%\end{eqnarray}
In these relations we have defined
 ${\cal P}_1=\sqrt{-g} E^r_\theta$, 
${\cal P}_2=\sqrt{-g}r(E^r_r-E^\theta_\theta)/2$
and $d\bar r=dr/{r }$.
Therefore, the weighted constraints  $E_\theta^r$ and $E_r^r-E_\theta^\theta$
still satisfy Laplace equations in $(\bar r,\theta)$ variables.
Then, they
are fulfilled, when one of them is satisfied on the boundary and the other
at a single point
\cite{Wiseman:2002zc}.
From the boundary  conditions above,
it turns out that this is the case,
 $i.e.$ the numerical scheme is self-consistent.
 
Our numerical treatment can be summarized as follows.
The first step is to introduce a new radial variable  
$$x\equiv \frac{r}{c+x} \ ,$$ 
which maps the semi--infinite region $[0,\infty)$ to the finite region $[0,1]$,
$c$ being an appropriate constant, which is chosen to be one for most of the solutions.
The resulting equations are solved by
using a fourth order
finite difference scheme.
The system of four equations is discretised on a grid with
$N_r\times N_\theta$ points,
where typically $N_r\sim 250$, $N_\theta \sim 50$.
The solutions were constructed by
using a Newton-Raphson method
and two different sofware packages,
the solver \textsc{fidisol/cadsol}
 \cite{schoen,schauder1992cadsol}
and the Intel \textsc{mkl pardiso} sparse direct solver \cite{pardiso,schenk},
together with \textsc{cedsol}\footnote{Complex Equations -- Simple
Domain partial differential equations SOLver is a C++ package
being developed by I. Perapechka.
} library.
The results are in very good agreement,
 the typical
errors being of order of $10^{-4}$.
The compilation of the numerical output is done by using the software \textsc{mathematica}.

%%%%%%%%%%%%%%%%%%%%%%%%%%%%%%%%%%%%%%%%%%%%%%%%%%%%%%%%%%%%%%%%%
%\subsection{Other cases: the spherically symmetric sector }
%\label{sec_spherical}
%%%%%%%%%%%%%%%%%%%%%%%%%%%%%%%%%%%%%%%%%%%%%%%%%%%%%%%%%%%%%%%%%
  
Let us close this section by remarking that
other solutions can be studied within 
 the
ansatz (\ref{ansatz}) and the approach described above.
The simplest ones are
 the spherically symmetric boson stars,
in which case
$F_2=F_1$, with
$F_0,F_1,\phi$ being functions of $r$ only.
The boundary conditions in this case are similar to those 
displayed above, except 
that the scalar field does not vanish
at $r=0$ and in the equatorial plane.

%%%%%%%%%%%%%%%%%%%%%%%%%%%%%%%%%%%%%%%%%%%%%%%%%%%%%%%%
\section{DBSs solutions }
\label{sec_res}
%%%%%%%%%%%%%%%%%%%%%%%%%%%%%%%%%%%%%%%%%%%%%%%%%%%%%%%%

\subsection{Domain and illustrative solutions}

The numerical results confirm the existence of solutions
of eqs. (\ref{eq-F0})-(\ref{Eq12})
with the boundary conditions  (\ref{bc})
representing DBSs.
The individual solutions do not exhibit any sign of pathological behaviour as discussed below;
in particular, no conical singularities
are detected at the level of numerical accuracy,
while the Ricci and Kretschmann scalars are finite everywhere.

\begin{figure}[ht!]
%\lbfig{rhfar}
\begin{center}
\includegraphics[height=.39\textwidth, angle =0 ]{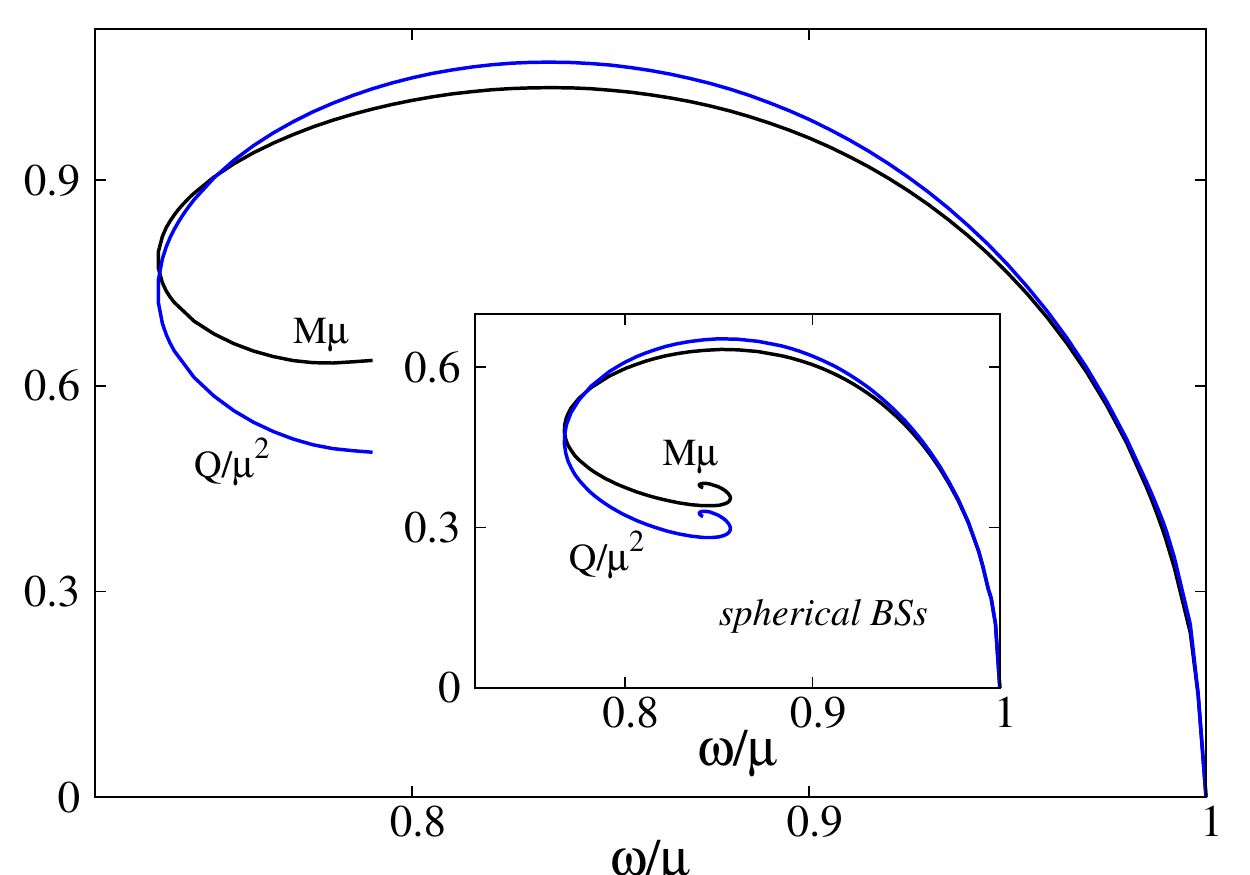} 
\end{center}
\caption{ 
The mass and Noether charge are shown $vs.$ scalar field frequency for the DBSs.
The inset shows the corresponding  picture for spherically symmetric, fundamental BSs~\cite{Herdeiro:2017fhv}, as a comparison.
}
\label{wM}
\end{figure}

The  ADM mass-frequency together with the  Noether charge-frequency diagrams are displayed in Figure \ref{wM}.
%
 %%%%%%%%%%%%%%%%%%%%%%%%%%%%%%%%%%%%%%%%%%%%%%%%%%%%%%%%%%%%%%%
%%%%%%%%%%%%%%%%%%%%%%%%%%%%%%%%%%%%%%%%%%%%%%%%%%%%%%%%%%%%%%%%
One notices that
the solutions exist for $  \omega <\mu$,
which is the usual bound state condition.
 As we decrease the frequency, the mass reaches monotonically the maximum value
$M_{\rm max} \mu \simeq  1.035$, for $\omega/ \mu \simeq  0.835 $.
The maximal value of $Q$ is also approached for the same solution, with
 $Q_{\rm max}\mu^2\simeq 1.072$.
Further decreasing $\omega$ one finds a minimal frequency
 $\omega_{\rm min} \sim 0.736 \mu$,
 below which no DBS solutions are found.
Instead,  a secondary branch of solutions occurs,
with $\omega$ increasing.
The DBS curve then seems to spiral,
 and
it is likely that the  
the picture familiar from the
spherically symmetric case
(see the inset in Figure \ref{wM})
 is recovered also in the dipole case.
Then we expect that
 the $(\omega,M)$ and $(\omega, Q)$
curves would  spiral towards a central region of the diagram, wherein the numerics become increasingly challenging.

As seen in Figure \ref{wM}, 
along the fundamental branch (with $\omega_{\rm min}<\omega<\mu$),
the ratio $\mu Q/M$ is greater than unity down to
a critical frequency
$\omega_c \sim 0.752 \mu$.
 
%%%%%%%%%%%%%%%%%%%%%%%%%%%%%%%%%%%%%%%%%%%%%%%%%%
\begin{figure}[h!]
\centering
\includegraphics[height=1.69in]{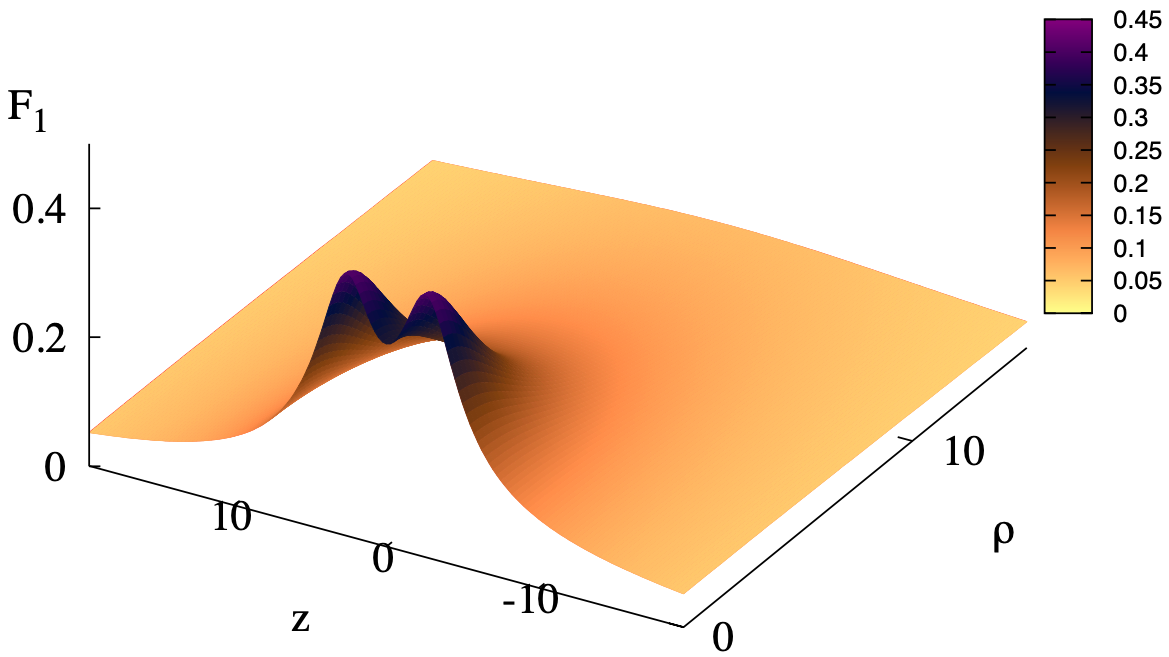}\ \ \ \ \ 
\includegraphics[height=1.72in]{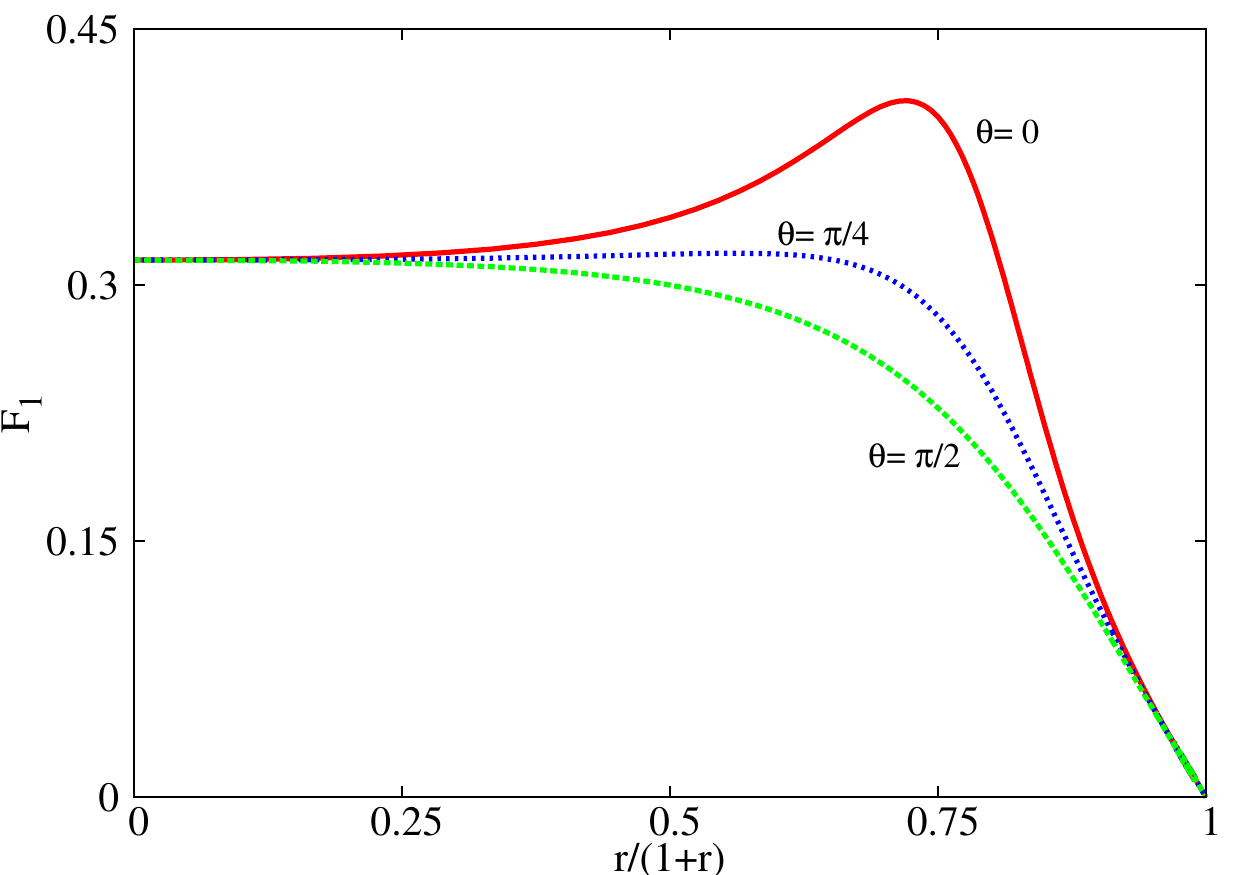}\\
\includegraphics[height=1.69in]{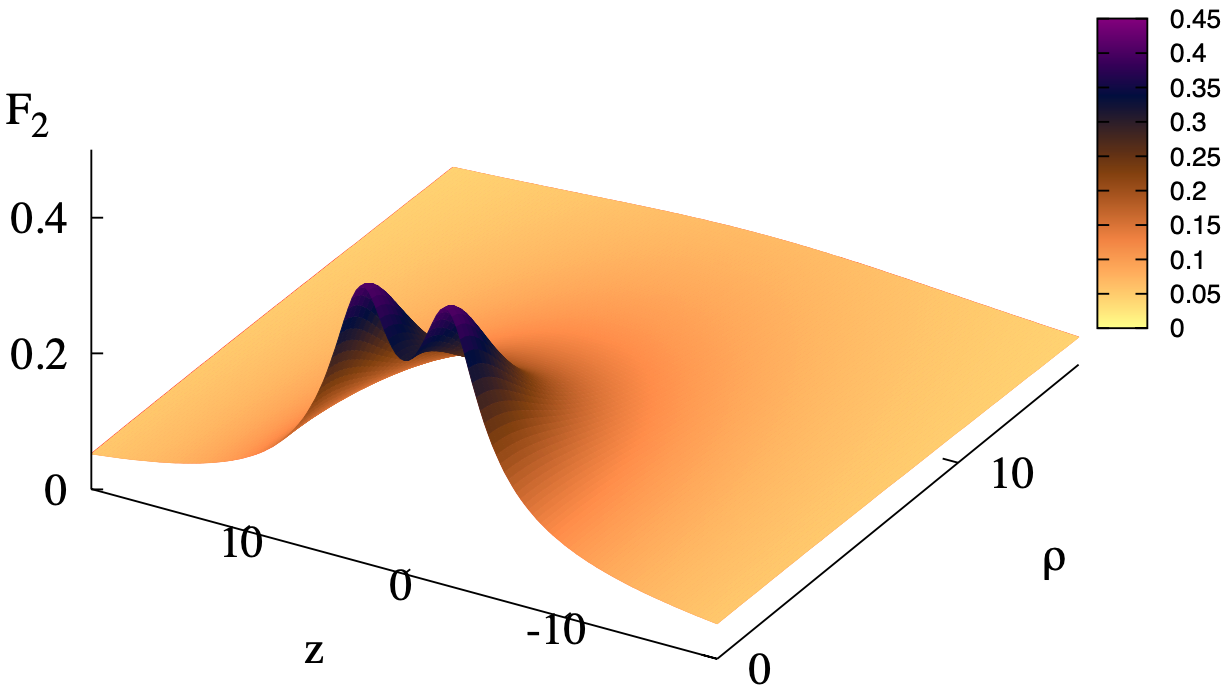}\  \ \ \ \ 
\includegraphics[height=1.72in]{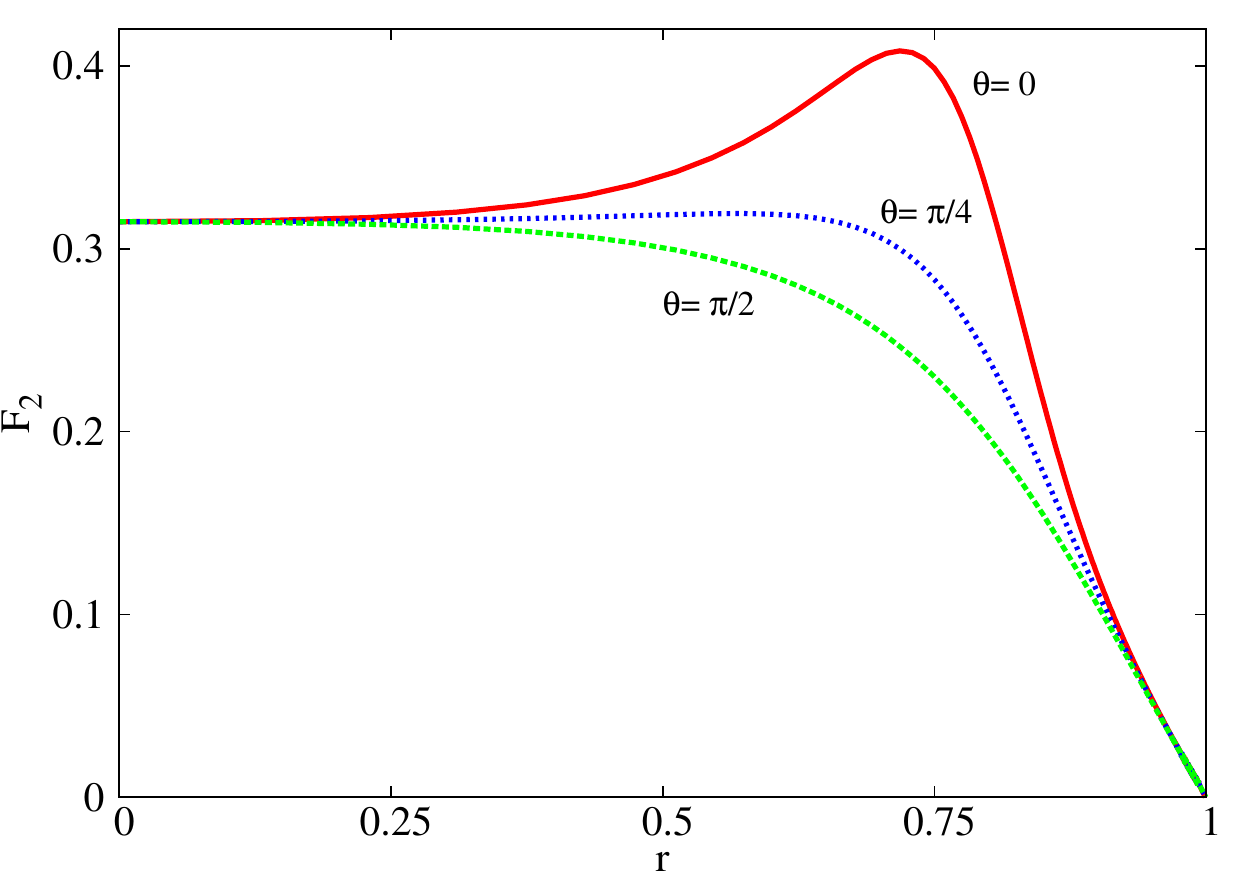}\\
\includegraphics[height=1.94in]{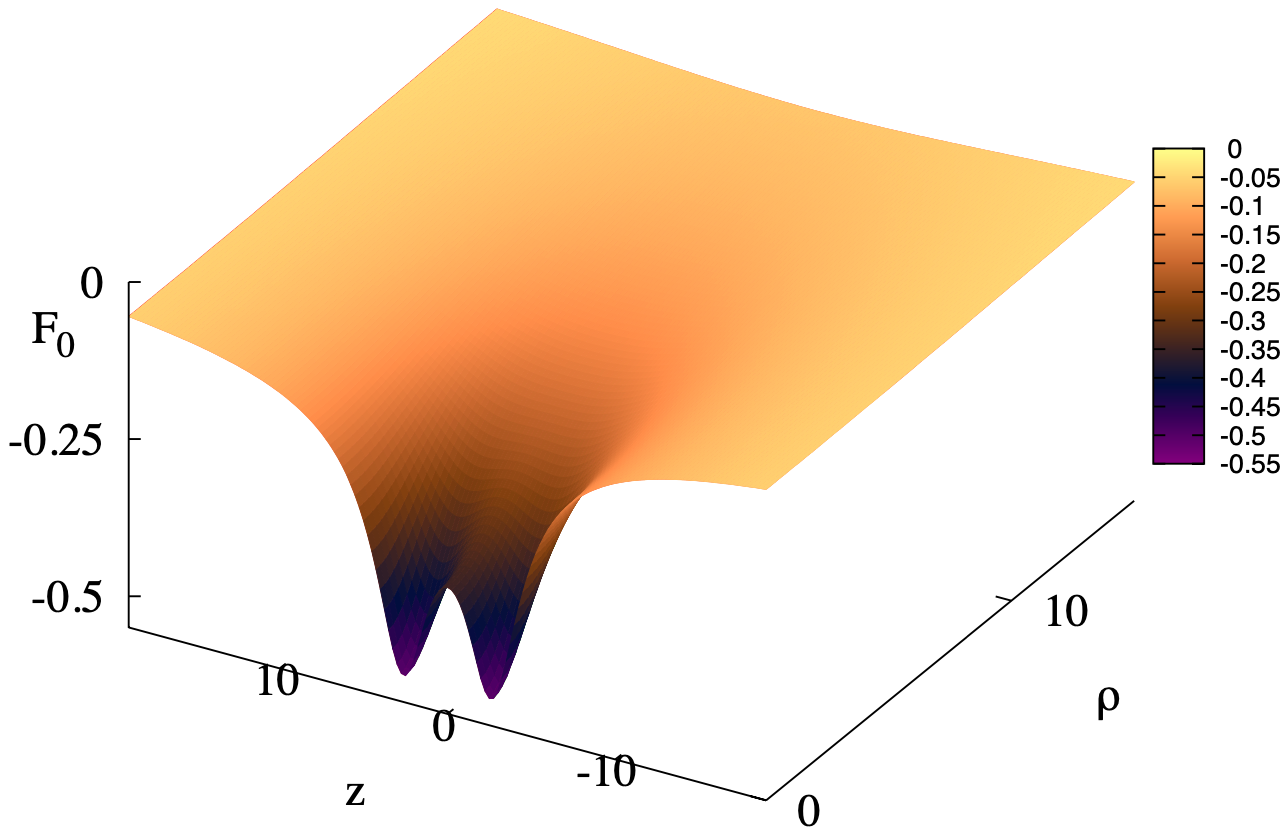}\  \ \ \ \ 
\includegraphics[height=1.72in]{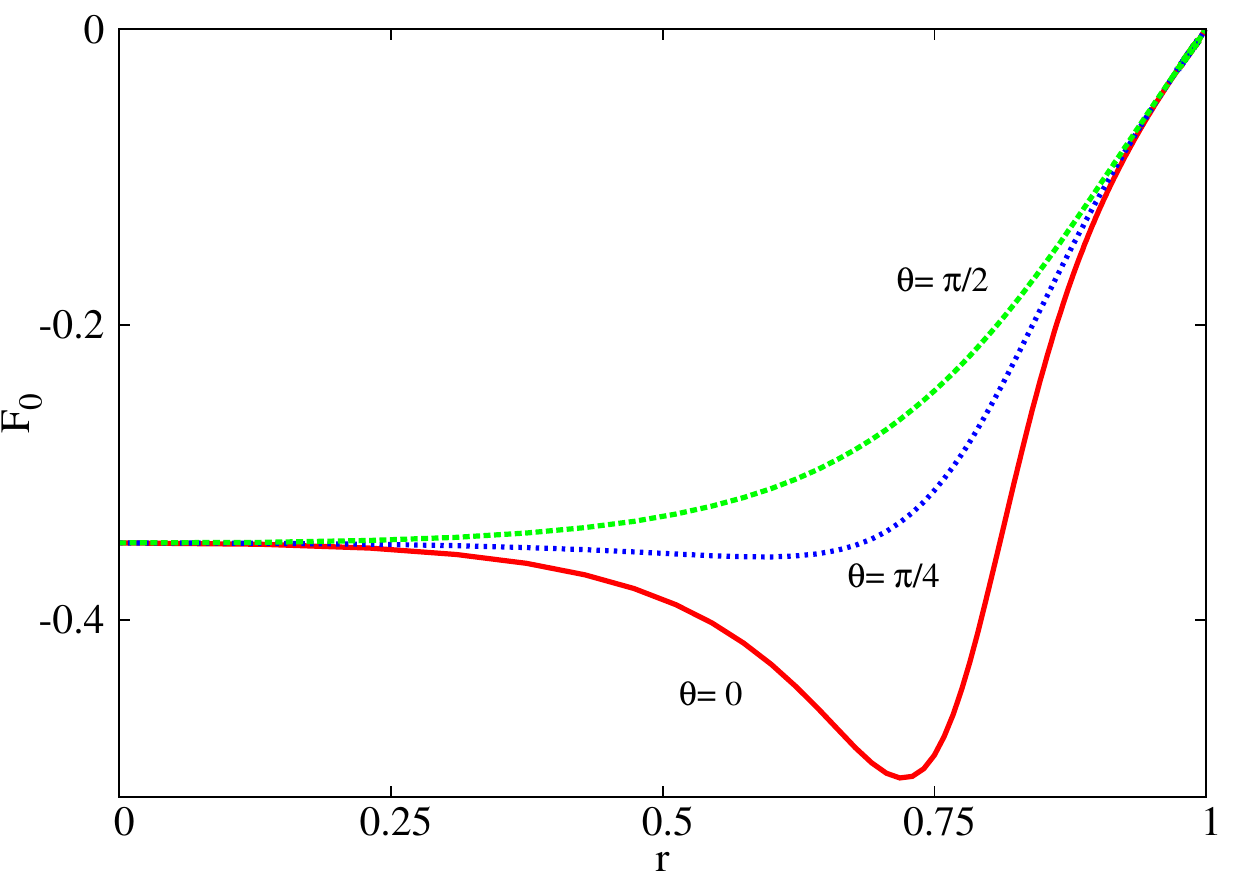}\\
\includegraphics[height=1.69in]{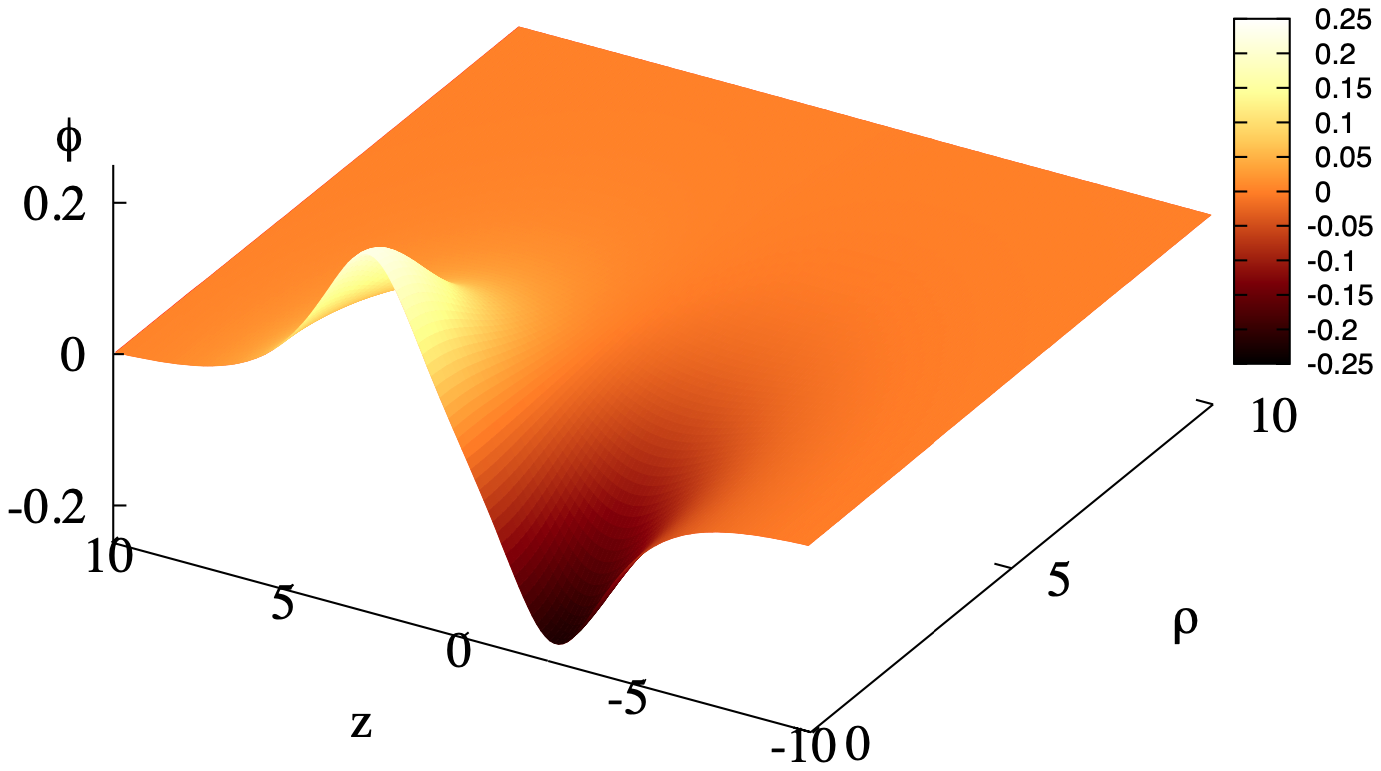}\ \ \ \ \ 
\includegraphics[height=1.72in]{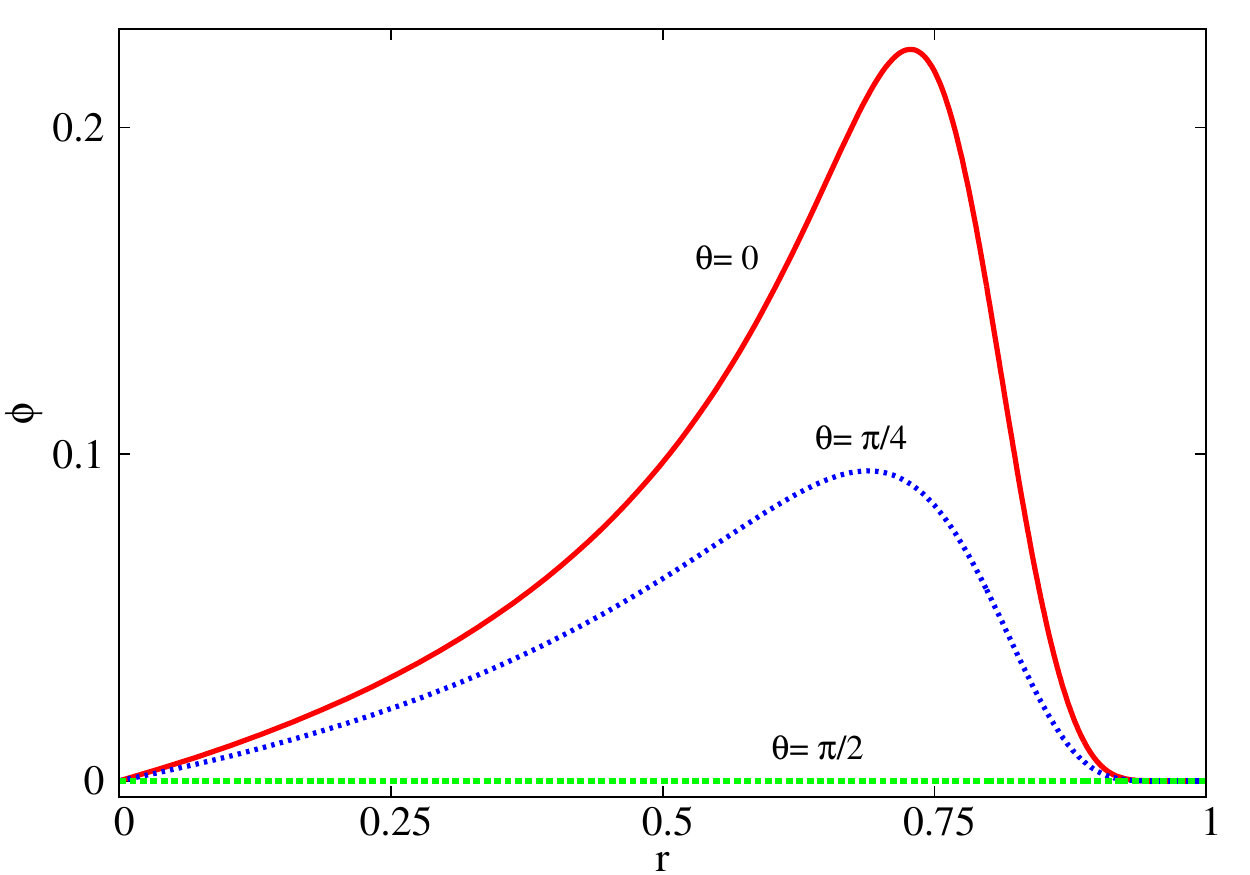}
\\
\caption{The metric functions $F_i$ 
and the scalar field $\phi$ are shown for   
the illustrative solution~(\ref{illsol}).} 
\label{sol3D2D}
\end{figure}
%%%%%%%%%%%%%%%%%%%%%%%%%%%%%%%%%%%%%%%%%%%%%%%%%%	
A standard energetic argument stating that the BS has more energy than a system of $Q$ free bosons, 
implies that the set of DBSs beyond that point -- $i.e.$ for $\omega<\omega_c$ along the first branch and on the whole second branch -- are unstable against fission, since the system prefers to relax into free bosons. Whereas, this does not mean that this is the true fate of such stars, it establishes they are unstable.

\begin{figure}[h!]
\centering
\includegraphics[height=1.69in]{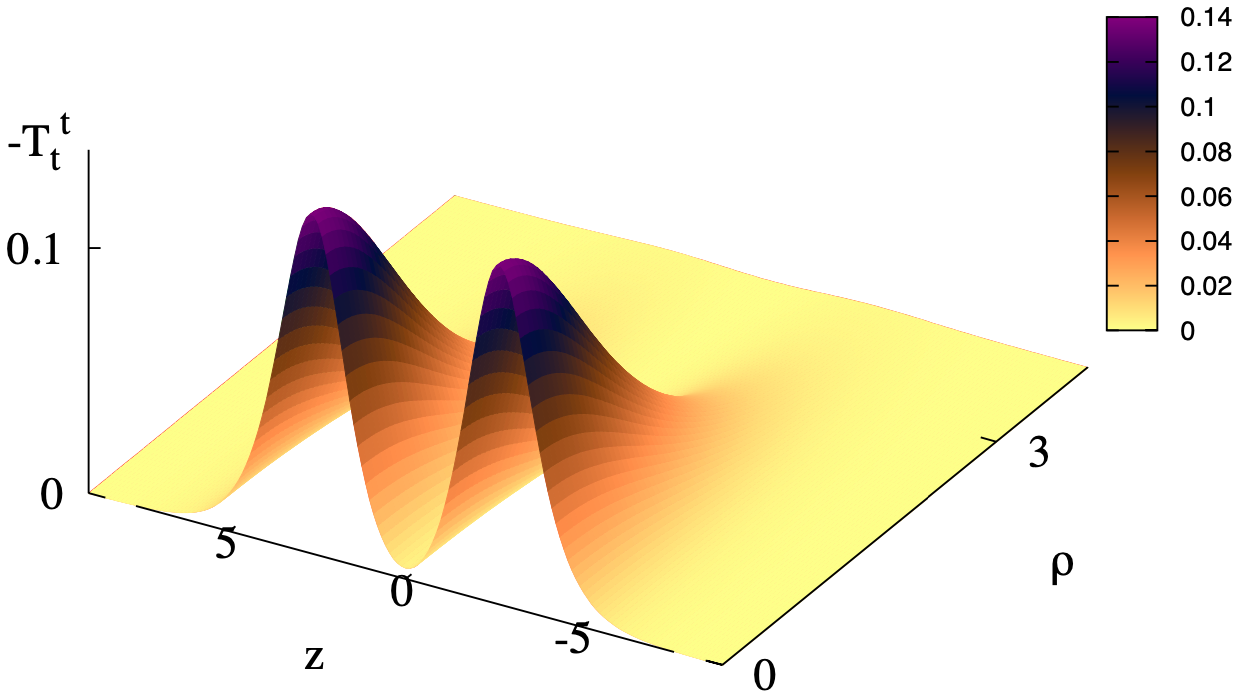}\ \ \ \ \ 
\includegraphics[height=1.72in]{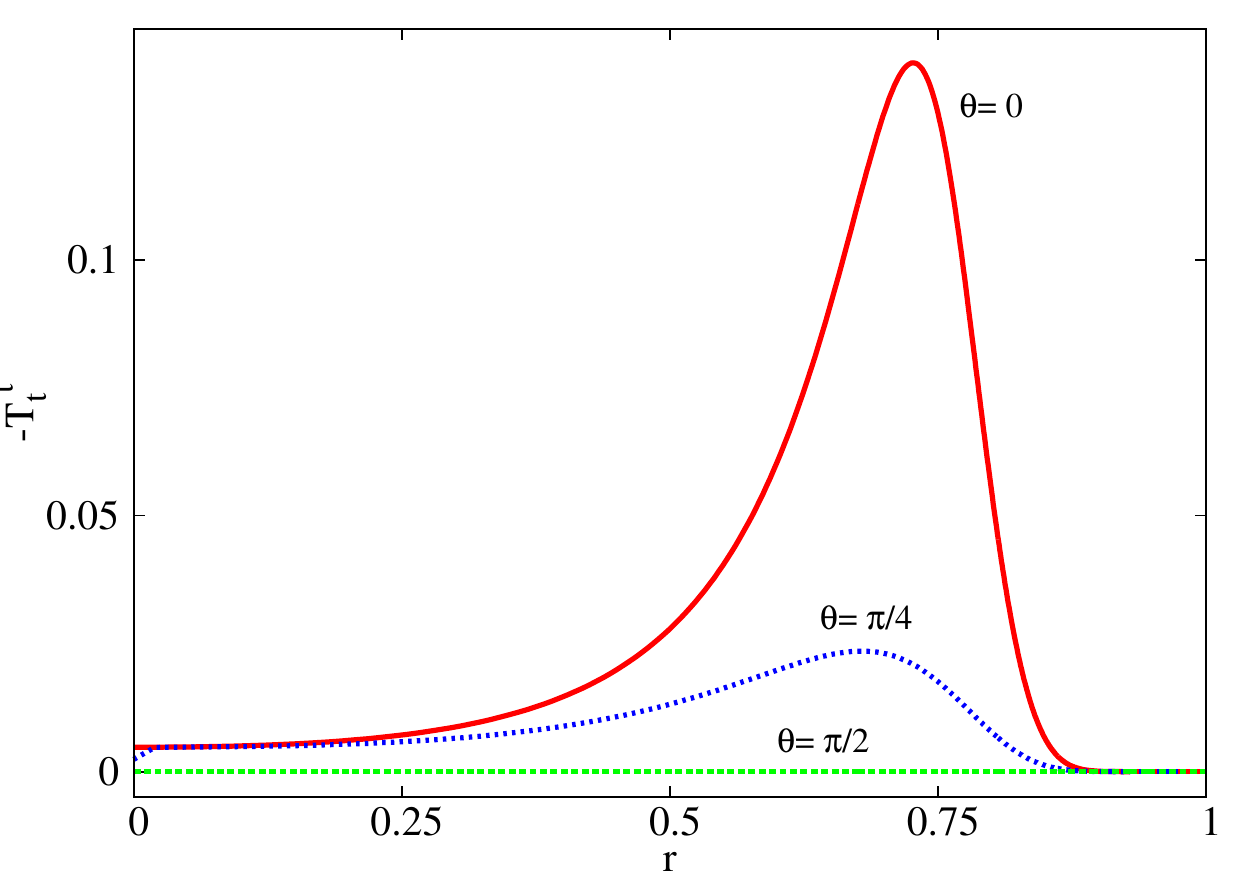}\\
\includegraphics[height=1.69in]{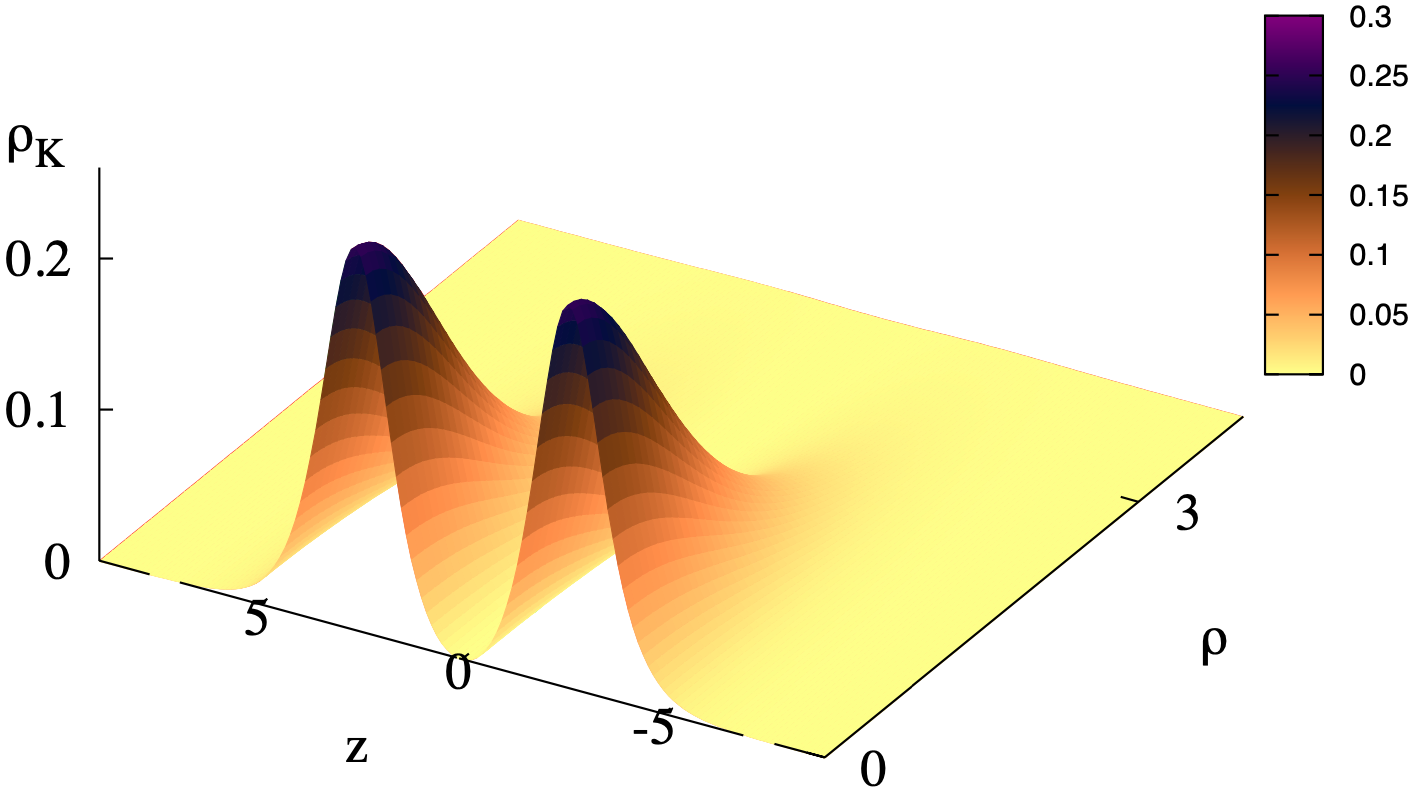}\  \ \ \ \ 
\includegraphics[height=1.72in]{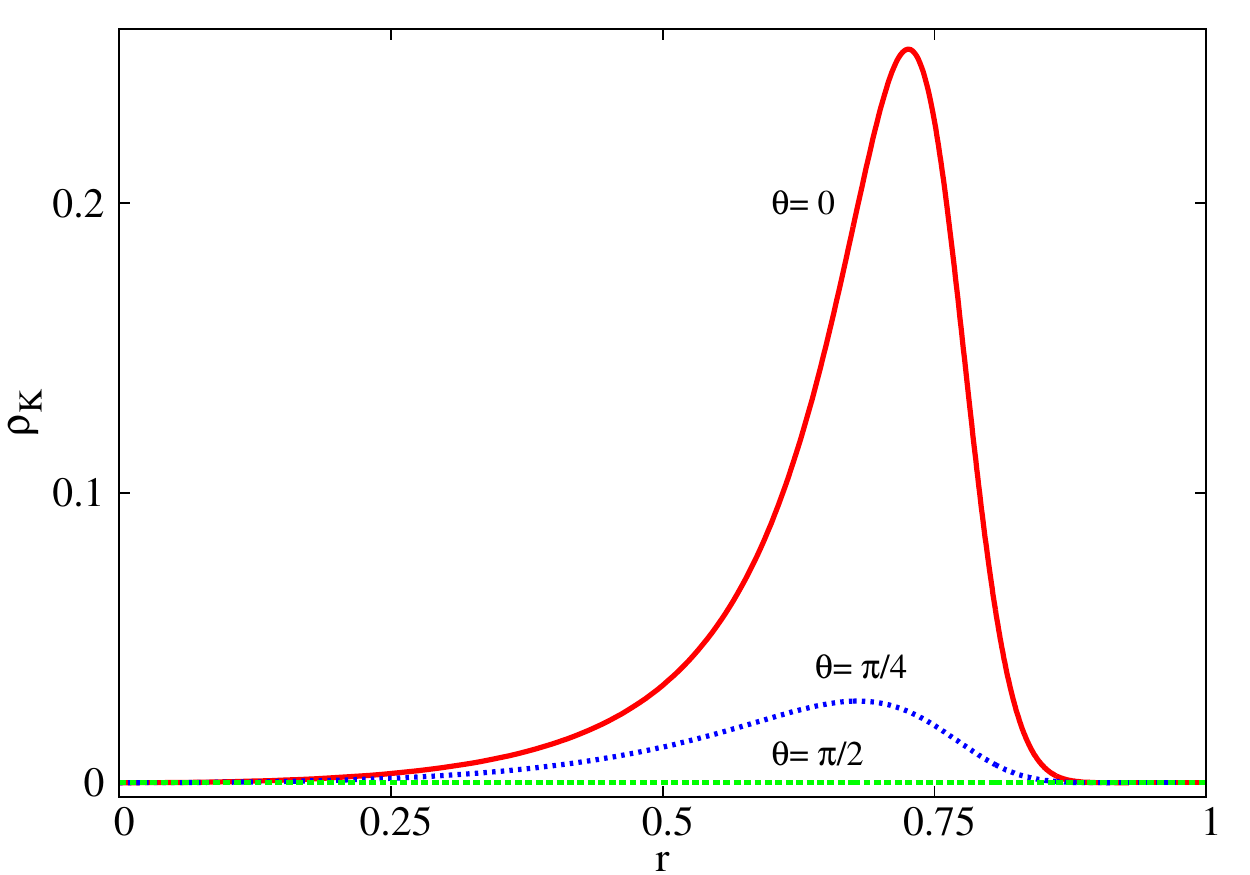}\\
\includegraphics[height=1.69in]{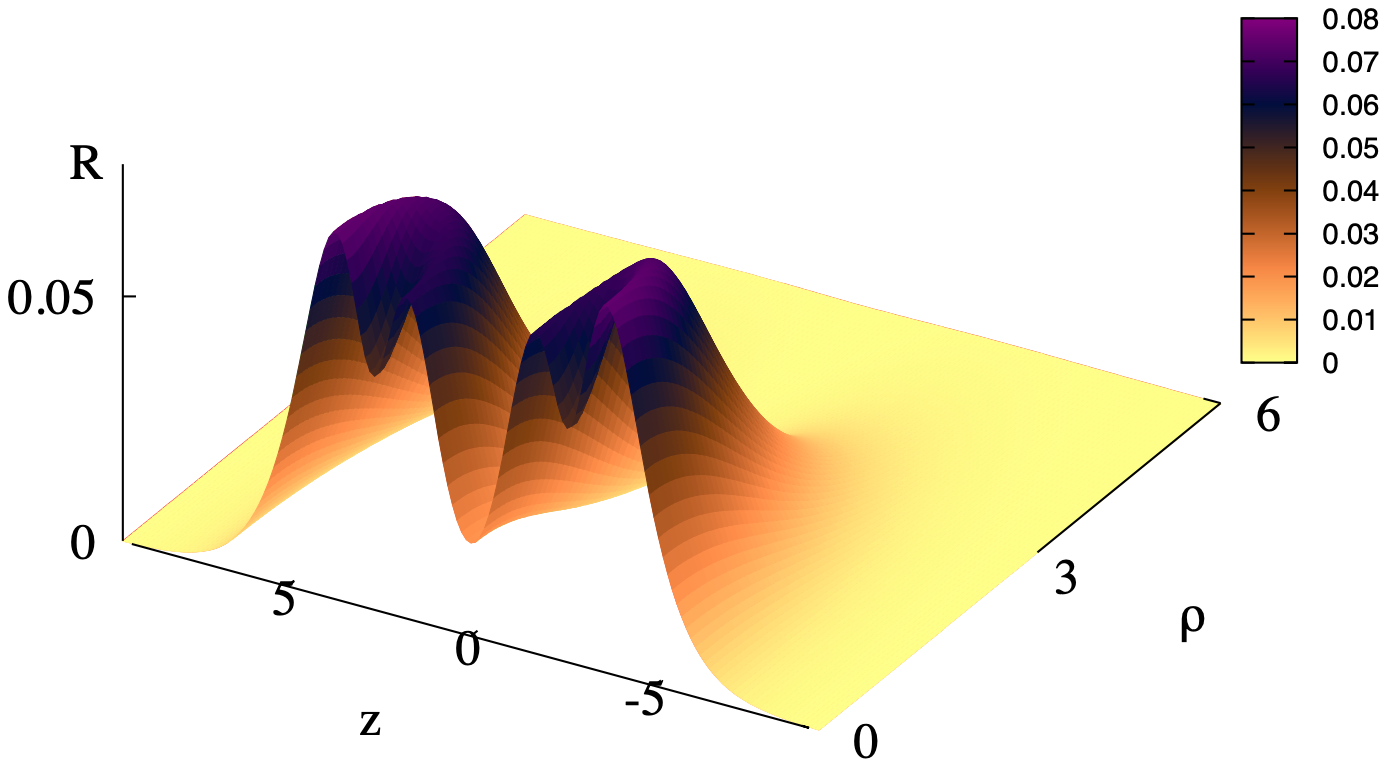}\  \ \ \ \ 
\includegraphics[height=1.72in]{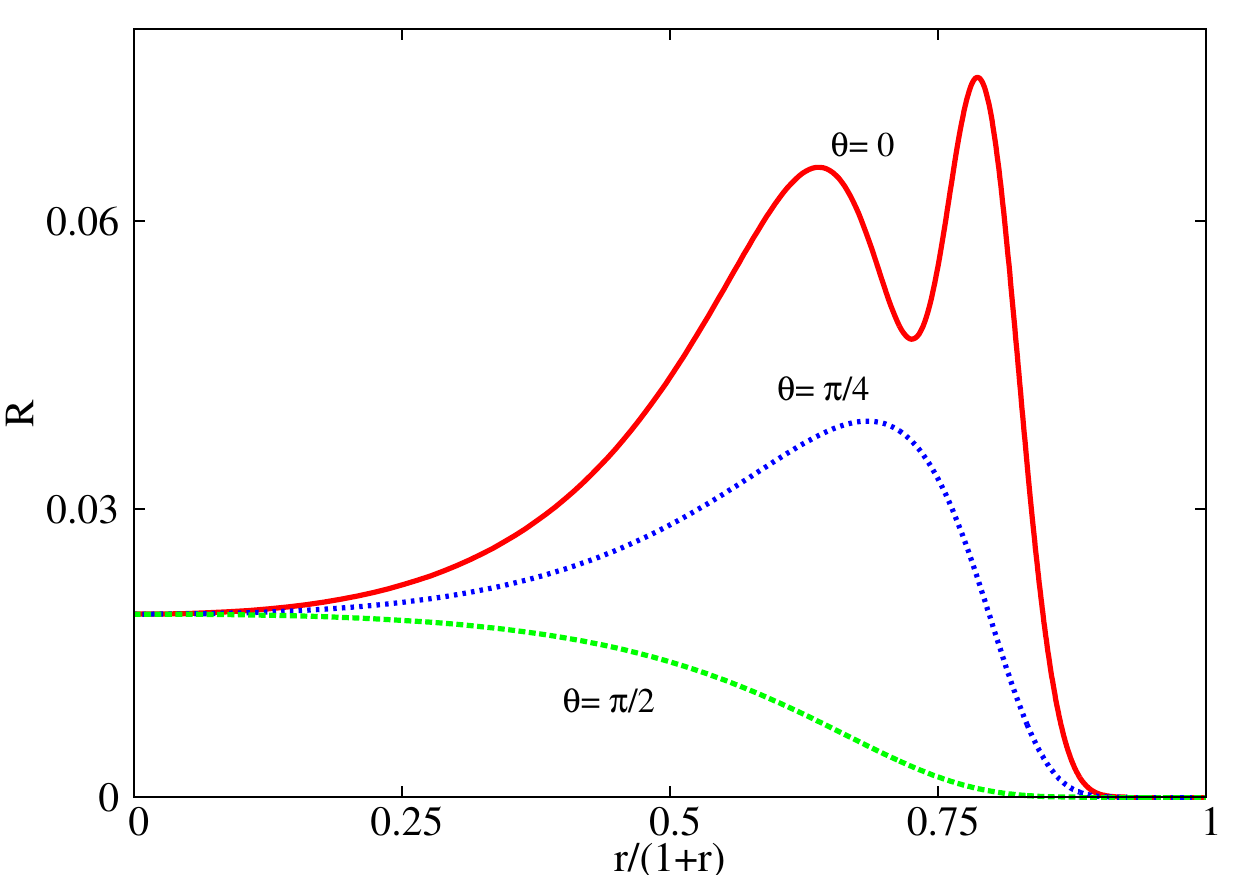}\\
\includegraphics[height=1.69in]{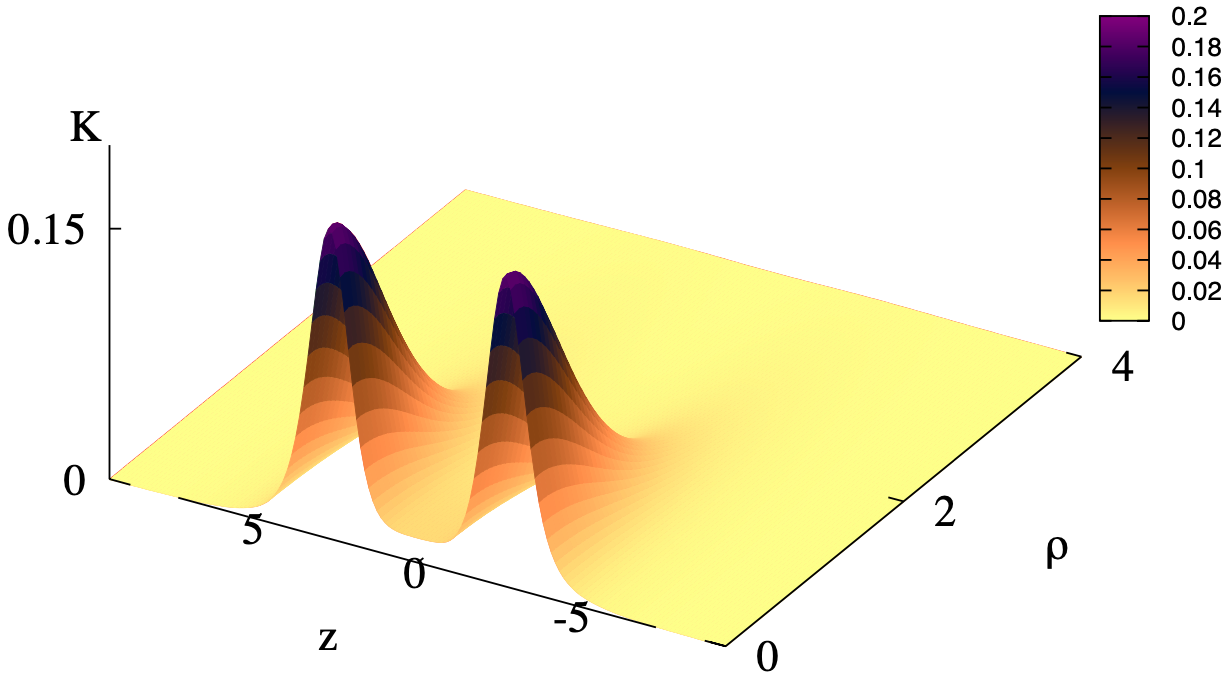}\  \ \ \ \ 
\includegraphics[height=1.72in]{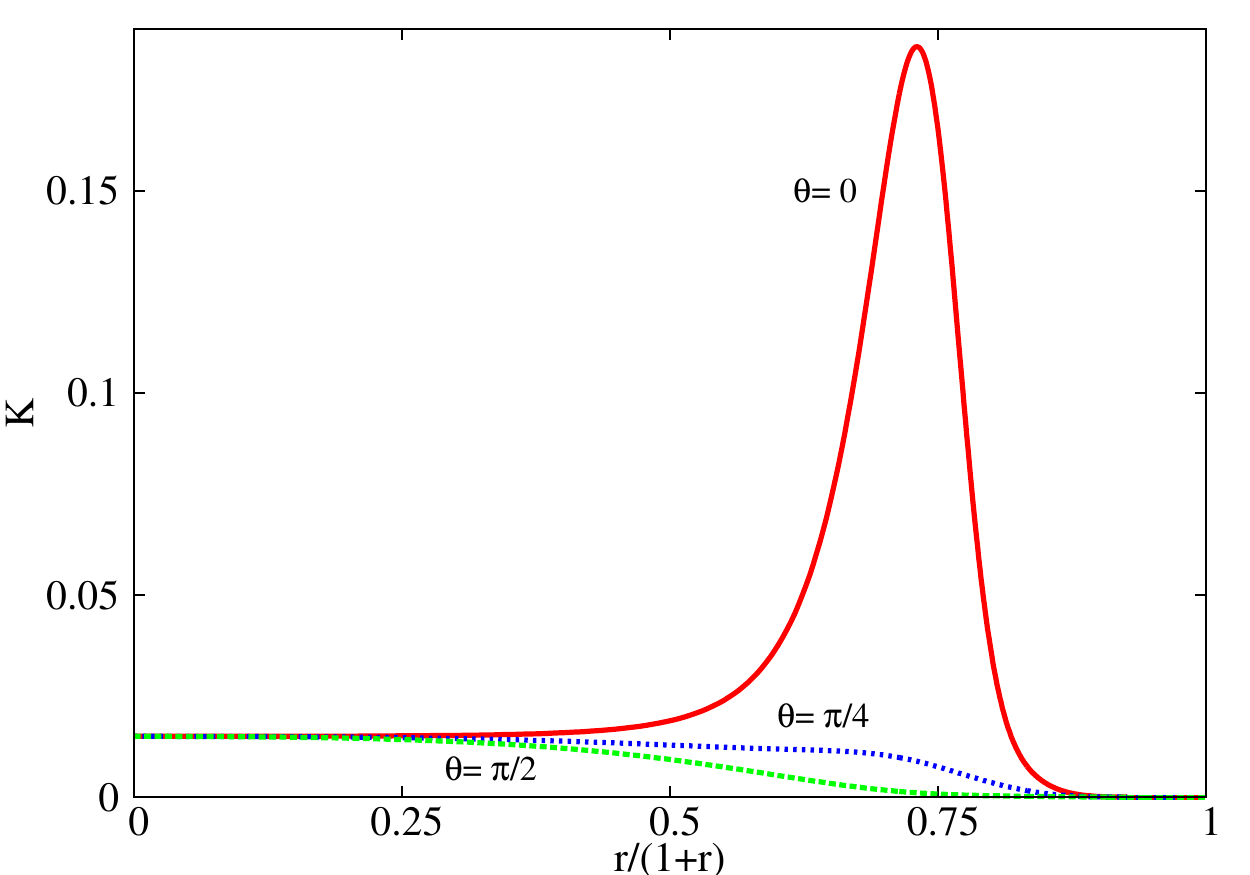}
\\
\caption{The component $T_t^t$
of the energy-momentum tensor, 
the Komar energy density
$\rho_K=T-2T_t^t$  and the Ricci and Kretschmann scalars are shown for the illustrative solution~(\ref{illsol}).
} 
\label{solRKrho}
\end{figure}

In Figures~\ref{sol3D2D} and \ref{solRKrho} 
we display various relevant functions and physical quantities   for a 
typical, illustrative solution with 
\begin{equation}
\label{illsol}
\omega/\mu=0.8 \ , \qquad M\mu =1.0165 \ , \qquad  Q\mu^2=1.049 \ , \qquad L_z \mu=5.361 \ .
\end{equation}

%%%%%%%%%%%%%%%%%%%%%%%%%%%%%%%%%%%%%%%%%%%%%%%%%%
%%%%%%%%%%%%%%%%%%%%%%%%%%%%%%%%%%%%%%%%%%%%%%%%%%	

The left columns displays 3D plots (in cylindrical coordinates), 
whereas the
right columnn shows 2D plots of the same 
functions in terms of the radial variable for three different
angular coordinates
$\theta=0$ (red solid line),
$\theta=\pi/4$ (blue dotted line)
and
$\theta=\pi/2$ (green dotted line).
A compactified radial coordinate is used, such
such that the asymptotic  behaviour of the functions
becomes transparent. Note in particular from  Figure~\ref{sol3D2D} (bottom panel) that the scalar field is parity odd.

%%%%%%%%%%%%%%%%%%%%%%%%%%%%%%%%%%%%%%%%%%%%%%%%%%%%%%%%%%%%%%%
\begin{figure}[ht!]
%\lbfig{rhfar}
\begin{center}
\includegraphics[height=.5\textwidth, angle =0 ]{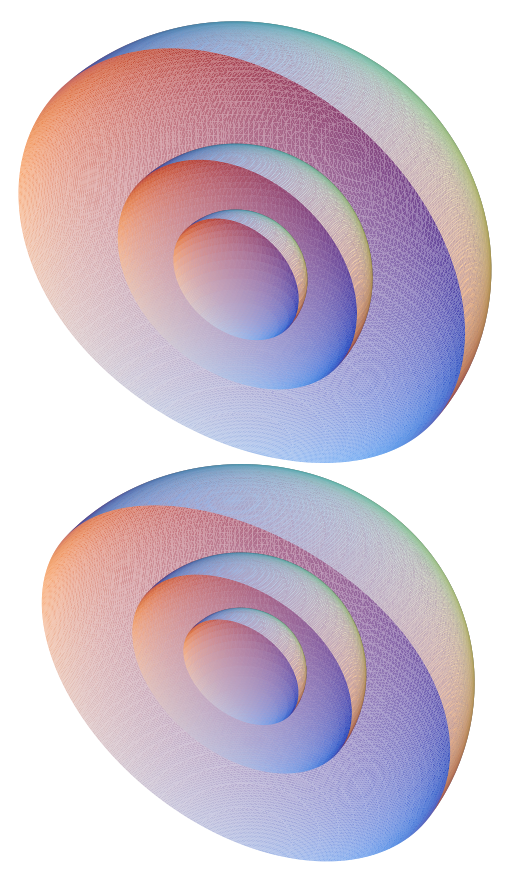} 
\end{center}
\caption{ 
 Surfaces of constant energy density, with $-T_t^t=0.1,0.05,0.01$, from the innermost to the outermost surface, 
are shown for the illustrative solution~(\ref{illsol}).
}
\label{isosurfaces}
\end{figure}
%%%%%%%%%%%%%%%%%%%%%%%%%%%%%%%%%%%%%%%%%%%%%%%  %%%%%%%%%%%%%%%%%%%%%%%%%%%%%%%%%%%%%%%%%%%%%%%%%%%%%%%%%%%%%%%
\begin{figure}[ht!]
%\lbfig{rhfar}
\begin{center}
\includegraphics[height=.39\textwidth, angle =0 ]{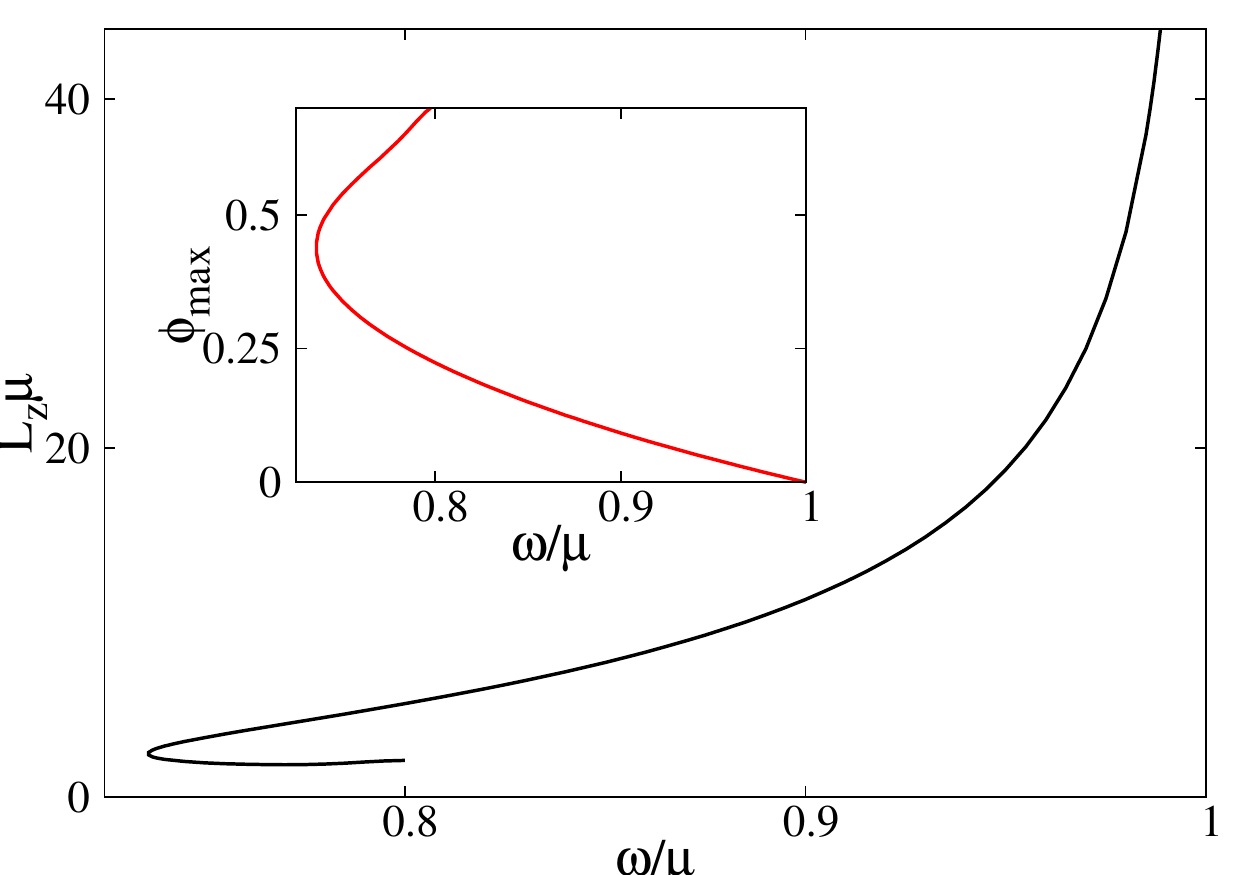} 
\end{center}
\caption{ 
The proper distance between the individual lumps and the  maximal value of the scalar field 
(inset) are
 shown 
as a function of scalar field frequency $\omega$ for the DBSs.
}
\label{Lm}
\end{figure}
%%%%%%%%%%%%%%%%%%%%%%%%%%%%%%%%%%%%%%%%%%%%%%%%%%%%%%%%%%%%%%%%

Figure \ref{isosurfaces} shows the morphology of the same illustrative solution by exhibiting the surfaces of constant energy density. DBSs always possess $two$  distinct components,
the energy of the scalar field being located around two distinct centers, 
located on the $z-$axis, at $z= \pm z_0$. 
This also corresponds to maximum of various other quantities like $|\phi|,R,K$ -- see Figures
\ref{sol3D2D} and \ref{solRKrho}.
Then a surface of constant energy density
(as given by the 
$T_t^t$ component of the energy-momentum tensor)
yields two spheroidal surfaces located on the symmetry axis at $z=\pm z_0$ --  Figure \ref{isosurfaces}.
Let us remark that the energy density (unlike the scalar field) does not vanish
at $r=0$: $T_t^t=-d_1^2 e^{-2 f_{10}}$, where $d_1,f_{10}$ are the constants in the small-$r$
expansion (\ref{small-r}).

 %%%%%%%%%%%%%%%%%%%%%%%%%%%%%%%%%%%%%%%%%%%%%%%%%%%%%%%%%%%%%%%
\begin{figure}[ht!]
%\lbfig{rhfar}
\begin{center}
\includegraphics[height=.39\textwidth, angle =0 ]{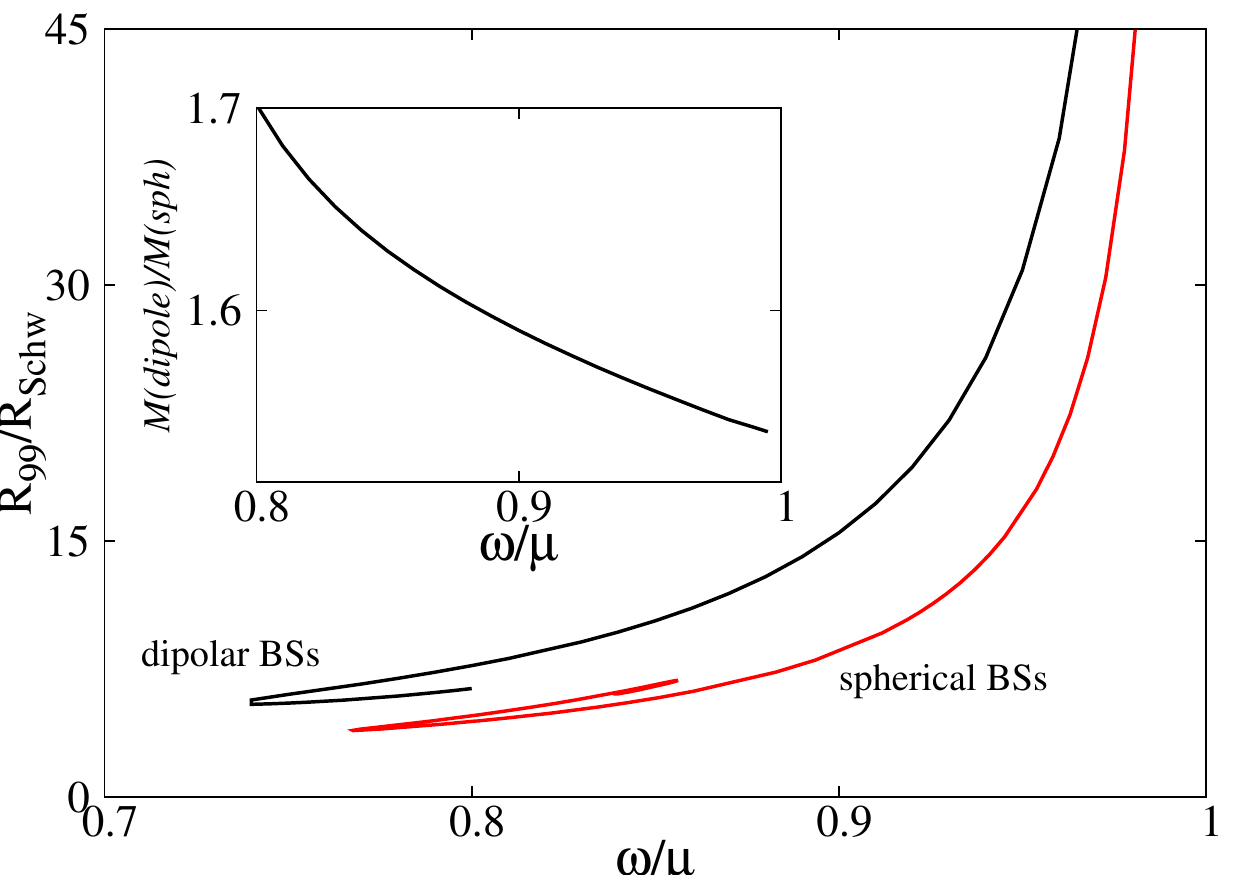} 
\end{center}
\caption{ 
Inverse compactness of DBSs.
The inset shows the ratio between the mass of a (fundamental branch) DBS and a spherical one,
both with the same frequency.
}
\label{comp}
\end{figure}
%%%%%%%%%%%%%%%%%%%%%%%%%%%%%%%%%%%%%%%%%%%%%%%%%%%%%%%%%%%%%%%%

In Figure \ref{Lm} we display the proper distance $L_z$, as
defined by eq. (\ref{Lz}), 
as a function of the frequency.
The two BSs become infinitely separated as $\omega\to \mu$; then,  
the distance between them decreases with $\omega$.
The minimal value of $L_z$ is approached on the secondary branch,
for $\omega/\mu \simeq 0.77$, with the proper distance increasing again after this point.
The maximal value of the scalar field, on the other hand, which is approached
on the $z$-axis, at $z=z_0$, always
increases along the spiral -- Figure \ref{Lm} (inset). The latter behaviour is analogous to that of single spherical BSs.

The
(inverse) compactness of DBSs is shown in Figure \ref{comp}.
Although these odd parity solutions are less compact than single, spherical BSs, 
along the fundamental branch,
the usual diagram for a spherical BS is recovered.
The picture changes, however,  along the secondary branch, where
we notice an increase in compactness relatively to the first branch, in contrast to the spherical case, where the compactness decreases in the second branch relatively to the first one.

As one can see in the inset of Figure \ref{comp},
for a given frequency,
the mass of a DBS is 
smaller than the mass of two spherical BSs with the same frequency. This is consistent with the DBSs being a bound state (with negative binding energy) of two individual BSs. 
Here we consider first branch solutions only,
where such a comparison is more meaningful, and recall that the spherical 
BSs
exist for $\omega_{\rm min}<\omega<\mu$,
with a maximal mass $M_{\rm max} \mu \sim 0.633$. Then, we may expect that the binding energy goes to zero and the system becomes better approximated by a linear superposition of two BSs as $\omega\rightarrow \mu$. Let us give evidence this is the case.

As $\omega \to \mu$,
the DBS is indeed a bound state of
two (largely separated) BSs, 
then a single BS would carry half of the total Noether charge,
\begin{eqnarray}
  Q_{\rm BS}=\frac{1}{2}Q_{\rm DBS} \ .
\end{eqnarray} 
This provides a range of $Q_{\rm BS}$
which indeed exists within the existing set of (spherical) BSs data.
Then, from the  numerical single BSs data
we compute their mass
$M_{\rm BS}$
corresponding to that $Q_{\rm BS}$.
It turns out that, as expected,
\begin{eqnarray}
1-\frac{2M_{\rm BS}}{M_{\rm DBS}} \simeq 10^{-3} \ .
\end{eqnarray} 
The corresponding individual spherical BSs and DBS
possess  slightly different 
frequencies, which are expected to converge as $\omega \to \mu$. We can support this expectation considering  
  a fitting of the
existing data with $\omega \to \mu$.
For DBSs one finds
\begin{eqnarray}
\label{relDBS}
 M_{\rm DBS} \simeq 3.80046
\sqrt{ 1-\frac{\omega }{\mu }} \ ,
\end{eqnarray} 
while the corresponding relation for BSs
reads
\cite{Friedberg:1986tp}
\begin{eqnarray}
\label{relsBS}
 M_{\rm BS} \simeq  2.47864  \sqrt{ 1-\frac{\omega }{\mu }} \ ,~~
\end{eqnarray}
which agrees very well with our numerical results.

Then 
let  us assume $M_{\rm DBS} (\omega_{(\rm DBS)})=2  M_{\rm BS}(\omega_{(\rm BS)}) =M$
and invert the relations
(\ref{relDBS}),
(\ref{relsBS})
to compute
the ratio
$\omega_{\rm (DBS)}/\omega_{\rm (BS)}$.
For small $M$, this gives
\begin{eqnarray}
 \frac{\omega_{\rm (DBS)}}{\omega_{\rm (BS)}}=1-0.028543 M^2+\dots \ ,
\end{eqnarray}
which is the expected result,
with
 $M_{\rm DBS} (\omega )=2  M_{\rm BS}(\omega )$ 
as $M\to 0$.
We also remark that a fitting of the  DBS data found for 
$\omega \to \mu$
gives the following relation
\begin{eqnarray}
M=\mu Q(1-c_0 \mu^2 Q^2) \ ,~~{\rm with}~~~c_0=0.0230503 \ .
\end{eqnarray} 
 This agrees well with  the result 
$c_0=0.02298$
in~\cite{Schupp:1995dy}
obtained within a Newtonian approximation.

%%%%%%%%%%%%%%%%%%%%%%%%%%%%%%%%%%%%%%%%%%%%%%%%%%%%%%%%%%%%%%%%%
\subsection{Comparison with the simple effective model.}
%
%%%%%%%%%%%%%%%%%%%%%%%%%%%%%%%%%%%%%%%%%%%%%%%%%%%%%%%%%%%%%%%%%%
We shall now show how the simple model based on the interaction energy~(\ref{emodel}) for two point masses in flat spacetime,  captures the variation of the distance $L_z$ between the two centres of the DBSs in the fully non-linear solutions.

To allow the comparison, we take  the interaction energy~(\ref{emodel}) with $M\rightarrow M/2$ and $Q\rightarrow Q/2$, so that $M,Q$ can be compared with the ADM mass and Noether charge of the DBS, thinking of each component has having half of the mass/Noether charge. 
Then the total force between the two constituents ${\bf F}=-\nabla U$ has radial magnitude 
\begin{eqnarray}
\label{force}
F_g+F_s=-\frac{M^2}{4r^2}+g\frac{Q^2}{4r^2}(1+\alpha r)e^{-\alpha r} \ .
 \end{eqnarray} 
Equilibrium amounts to the force balance condition  
\begin{equation} 
F_g+F_s=0 \qquad {\rm for} \quad r=L_z \ .
\label{force-cons}
\end{equation}

We have tested the naive equilibrium condition (\ref{force-cons}) for the family of static DBSs reported above. For each DBS (specified by $\omega$, say) we use its $(M,Q,L_z)$ to compute $F_g$, $F_s$ above and determine $g,\alpha$ by enforcing (\ref{force-cons}).
\textit{A priori}, these parameters may vary significantly along the space of DBSs solutions.
Considering various pairs of frequencies
$(w_1,w_2)$
on the first branch of solutions -- where this simple mechanical model is more reasonable -- we always found a value of $g\simeq 1$ and $\alpha \simeq 0.01$,  with a few percent variation for both parameters.
Thus, in what follows we choose these values for $g,\alpha$.

\begin{figure}[h!]
\centering
\includegraphics[height=.39\textwidth, angle =0 ]{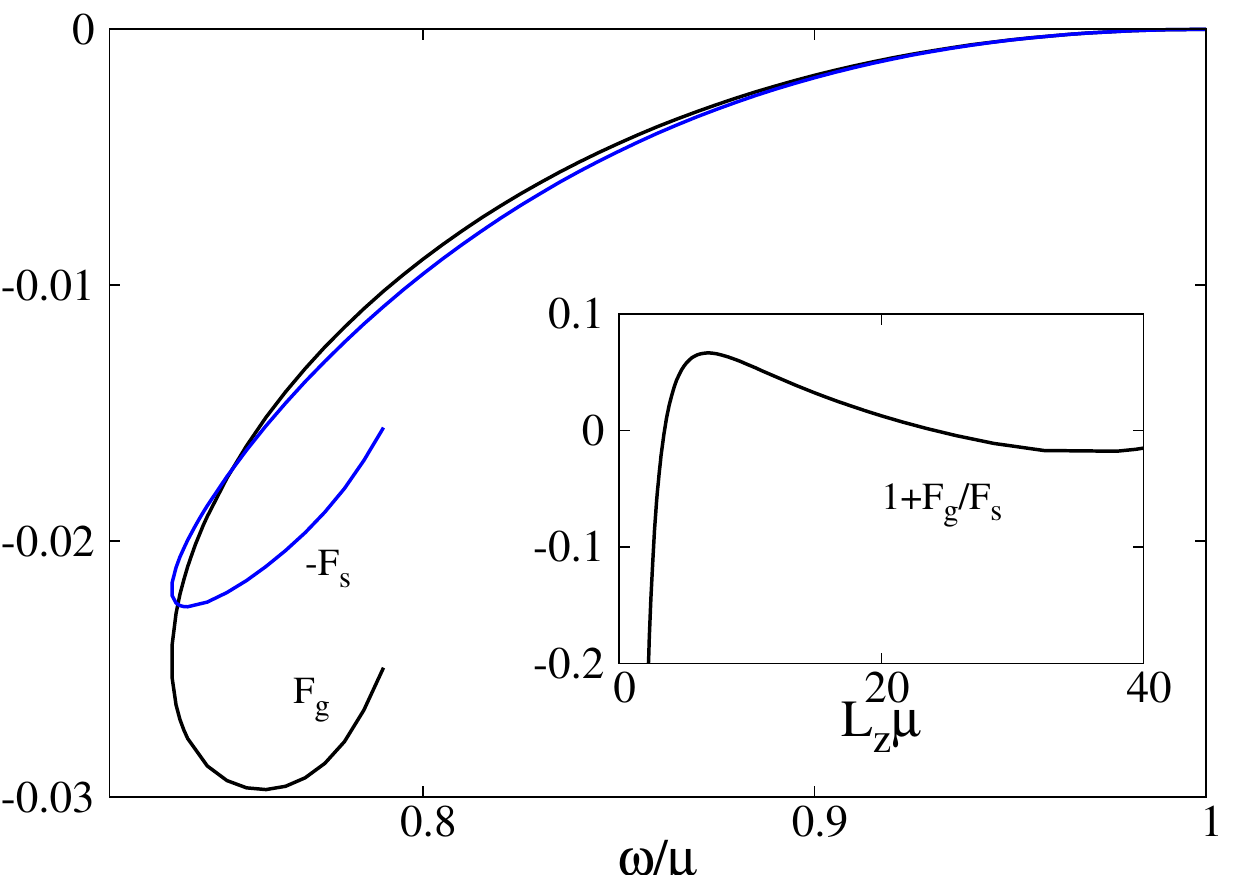} 
\\
\caption{The forces in the effective model --
$cf.$ (\ref{force}) -- 
and the relative difference between them.}
 
\label{fforce}
\end{figure}

In Figure \ref{fforce}
we show the individual forces $F_g$, $F_s$ 
as computed from (\ref{force}) (with $r=L_z$)
and also the relative difference between them (in the inset).
One observes that this simple model 
works relatively well for the set of solutions
between $\omega_{\rm max}=\mu$
and $\omega_{\rm min}$.
For example, the relative error found for 
the solution with the maximal mass and charge 
($\omega/\mu\simeq 0.835$)
is 
1+$F_g/F_s \sim 6\%$.
Likely, this agreement is partly a consequence of the fact that
the ratio
$Q/M$ is
 not too far from
unity for all solutions
between $\omega_{\rm max}$
and $\omega_{\rm min}$.

%%%%%%%%%%%%%%%%%%%%%%%%%%%%%%%%%%%%%%%%%%%%%%%%%%
%%%%%%%%%%%%%%%%%%%%%%%%%%%%%%%%%%%%%%%%%

%%%%%%%%%%%%%%%%%%%%%%%%%%%%%%%%%%%%%%%%%%%%%%%%%%%%%%%%%%%%%%%%%
\subsection{The role of gravity: no flat spacetime   static $Q$-balls in equilibrium}
%
%%%%%%%%%%%%%%%%%%%%%%%%%%%%%%%%%%%%%%%%%%%%%%%%%%%%%%%%%%%%%%%%%%
The last subsection corroborates the interpretation of DBSs can as two (static) BSs in equilibrium, by virtue of a balance between their gravitational attraction and their scalar repulsion. We shall now put forward a different argument that emphasises the importance of the long range gravitational interaction, for the existence of DBSs.
  
Single BSs have a flat spacetime cousin: $Q$-balls~\cite{Coleman:1985ki}.  $Q$-balls are interpreted as emerging from a balance between the dispersiveness of the oscillating scalar field (amplitude) and the attractiveness of the self-interactions. When turning on gravity, however, the scalar self-interactions are dispensable; their role is replaced by another attractive, non-linear interaction. There is, however, an important difference: the scalar (gravitational) interaction is short (long) range.

This discussion suggests that whereas non-linear scalar and gravitational interactions may be interchanged for obtaining single energy lumps in equilibrium, they might not mimic one another for distant energy lumps in equilibrium. This leads to the conjecture that no flat spacetime multi-$Q$-balls in equilibrium exist. We shall now follow the arguments in  \cite{Gibbons:2010cr}, 
to prove this statement, for the case of two equilibrium $Q$-balls, in a non-gravitating model with a complex scalar field.
Thus, gravity is a crucial ingredient for the existence of equilibrium configurations of two such scalar lumps, a result which is independent of the precise  scalar field potential.

The starting point of the argument assumes 
a generic field theory model on 
 flat spacetime,
with its metric in
 cylindrical coordinates
\begin{eqnarray}
ds^2=-dt^2+d\rho^2+\rho^2 d\varphi^2+dz^2 \ .
\end{eqnarray}
One focuses on static
axially symmetric 
solitonic configurations,
with a conserved 
energy-momentum tensor,
\begin{eqnarray}
\label{r1}
\nabla_\alpha T^\alpha_\beta =\frac{1}{\sqrt{-g}}\frac{\partial}{\partial x^\alpha}(\sqrt{-g}T^\alpha_\beta)
-\frac{1}{2}\frac{\partial g_{\alpha\nu}}{\partial x^\beta}T^{\alpha\nu}=0~.
\end{eqnarray}
Then, the only nonzero components of the energy-momentum tensor are
 \begin{eqnarray}
T_{\rho \rho}(\rho,z),~
T_{z z}(\rho,z),~
T_{\rho z}(\rho,z),~
T_{\varphi \varphi}(\rho,z),~
T_{t t}(\rho,z) \ .
\end{eqnarray}

Next, one consider the $x^\beta=z$ component of eq. 
(\ref{r1})
and integrates it over the half-space $z\geq 0$.
This implies
the general relation
\begin{eqnarray}
\label{r2}
 \int_0^\infty dz \rho~ T_z^\rho \big|_{\rho=0}^{ \infty}
+ \int_0^\infty d\rho~ \rho  T_z^z \big|_{z=0}^{ \infty} =0 \ ,
\end{eqnarray}
which must be satisfied by any static soliton.
Now, we assume that  $T_z^\rho$ is finite at $\rho=0$
and vanishes faster than $1/\rho^2$ as $\rho \to \infty$, both necessary conditions for a finite energy, everywhere regular soliton. This implies that the first integral in (\ref{r2}) vanishes. Additionally, we assume  that $ T_z^z$ goes to zero as $z\to \infty$;
then (\ref{r2})
reduces to 
\begin{eqnarray}
\label{relzx}
 \int_0^\infty d \rho  ~ \rho T_z^z (\rho,z) \big|_{z=0} =0 \ .
\end{eqnarray}
This is a generic constraint that we shall now evaluate for  a complex scalar field
$\Phi$ model, with the Lagrangian density
\begin{eqnarray}
\label{Q1}
\mathcal{L}=
  -\frac{1}{2} g^{\alpha\beta}\left( \Phi_{, \, \alpha}^* \Phi_{, \, \beta} + \Phi _
{, \, \beta}^* \Phi _{, \, \alpha} \right) -U(|\Phi|^2) \ ,
\end{eqnarray}
where 
$U(|\Phi|^2)$
is the self-interaction potential, 
with 
$U(0)=0$.
The scalar field satisfies the KG equation
\begin{eqnarray}
\label{Q1n}
\Box \Phi=\frac{\partial U}{\partial |\Phi|^2}\Phi \ . 
\end{eqnarray}

For a $Q$-ball solution, one assumes a (static) scalar field ansatz
\begin{eqnarray}
\label{Q2}
\Phi=e^{-i\omega t}\phi(\rho,z) \ ,
\end{eqnarray}
which results in the following expression for $T_z^z$:
\begin{eqnarray}
\label{Q3}
T_z^z=-\phi_{,\rho}^2+\phi_{,z}^2+w^2 \phi^2 -U(\phi) \ .
\end{eqnarray}
Let us assume the existence of a
small-$z$ Taylor expansion
expansion of the scalar field
\begin{eqnarray} 
\phi(\rho, z)=\sum_{k\geq 0}\phi_k(\rho)z^k \ ,
\end{eqnarray}
such that
the integral 
(\ref{r2})
becomes
\begin{eqnarray}
\label{relns}
 \int_0^\infty \rho 
\Big[
\phi_{1}^2-\phi_{0,\rho}^2+w^2 \phi_0^2 -U(\phi_0)
\Big]d\rho  
=0 \ .
\end{eqnarray}
However, a odd-parity scalar field
necessarily
vanishes on the equatorial plane
\begin{eqnarray}
\label{Qx3}
\phi_0=0 \ , 
\end{eqnarray}
(or more general, $\phi_{2k}=0$)
such that (\ref{relns}) becomes
\begin{eqnarray}
\label{relnz}
 \int_0^\infty \rho  
\phi_{1}^2 \, d\rho 
=0 \ .
\end{eqnarray}
The only way to satisfy the above relation is to assume
$\phi_{1}(\rho)=0$.
However, from the KG
equation
(\ref{Q1n})
one finds that the higher order coefficients in (\ref{Q3}) 
vanish
order by order,
$\phi_{2k+1}=0$.
Thus, we conclude that no odd-parity static scalar field 
smooth solutions may exist, in particular, no $Q-$dipoles.

We remark that the above non-existence result does not apply to rotating configurations. Rotation introduces a dipole-dipole interaction that can be used to balance two spinning $Q$-balls in flat spacetime. Such solutions were briefly reported in~\cite{Volkov:2002aj} - see also~\cite{Kleihaus:2007vk}.
Also, 
it does not exclude the existence of
even-parity solutions, 
the simplest one being the spherically symmetric $Q$-balls.

 %%%%%%%%%%%%%%%%%%%%%%%%%%%%%%%%%%%%%%%%%%%
\section{Gravitational Lensing}
\label{sec-lensing}
 %%%%%%%%%%%%%%%%%%%%%%%%%%%%%%%%%%%%%%%%%%%

The direct observation by the Event Horizon Telescope (EHT) of the supermassive BH candidates M87* and SgrA* has transformed the study of strong gravitational lensing into an active research field. The study of lensing properties of compact objects serves as a diagnosis of their strong gravity imprints - see $e.g.$~\cite{Cunha:2018acu} for a review. 
In this section we explore numerically the lensing images of DBSs, and discuss how some lensing features might be shared generically with other dipolar mass distributions with no horizon.

The observation image of a compact object, for a given observer, can be simulated numerically via {\it backwards ray-tracing}. Under this procedure, each pixel of the synthetic image corresponds to a slightly different observation direction in the observer's local sky, and contains the information of the light ray received along that direction. The pixel information can be modeled by propagating null geodesics backwards in time, starting from the observation location along the detection direction, until a light source can be found along the geodesic path - see $e.g.$~\cite{Bohn:2014xxa,Cunha:2016bpi} for details. 

We shall consider two different scenarios, both containing optically opaque light sources, as well as an optically transparent medium between the observer and the light sources.
In the first scenario, we shall consider a very large (far-away) colored sphere as the light source, centered around the dipole center and containing both the DBS and the observer inside it, following~\cite{Bohn:2014xxa,Cunha:2015yba}. The sphere is divided into four colored quadrants (red, yellow, green, blue), superimposed with a black mesh. This academic setup neatly illustrates how different image pixels are mapped into the far-away sphere and was previously used for two-BH systems in~\cite{Bohn:2014xxa,Cunha:2018gql,Cunha:2018cof}.
In the second scenario, we shall consider a geometrically thin accretion disk in the equatorial plane, with an inner edge at the circumferential radius of $6M$, which corresponds to the location of the Innermost Stable Circular Orbit (ISCO) in the Schwarzschild spacetime. To further mimic the effect of a realistic accretion disk, the light emission displays a typical decay profile as one moves further away from the disk's inner edge. This setup was previously used in $e.g.$~\cite{Cunha:2019hzj,Herdeiro:2021lwl}.

\begin{figure}[ht!]
     \centering
     \begin{subfigure}[b]{0.28\textwidth}
         \centering
         \includegraphics[width=\textwidth]{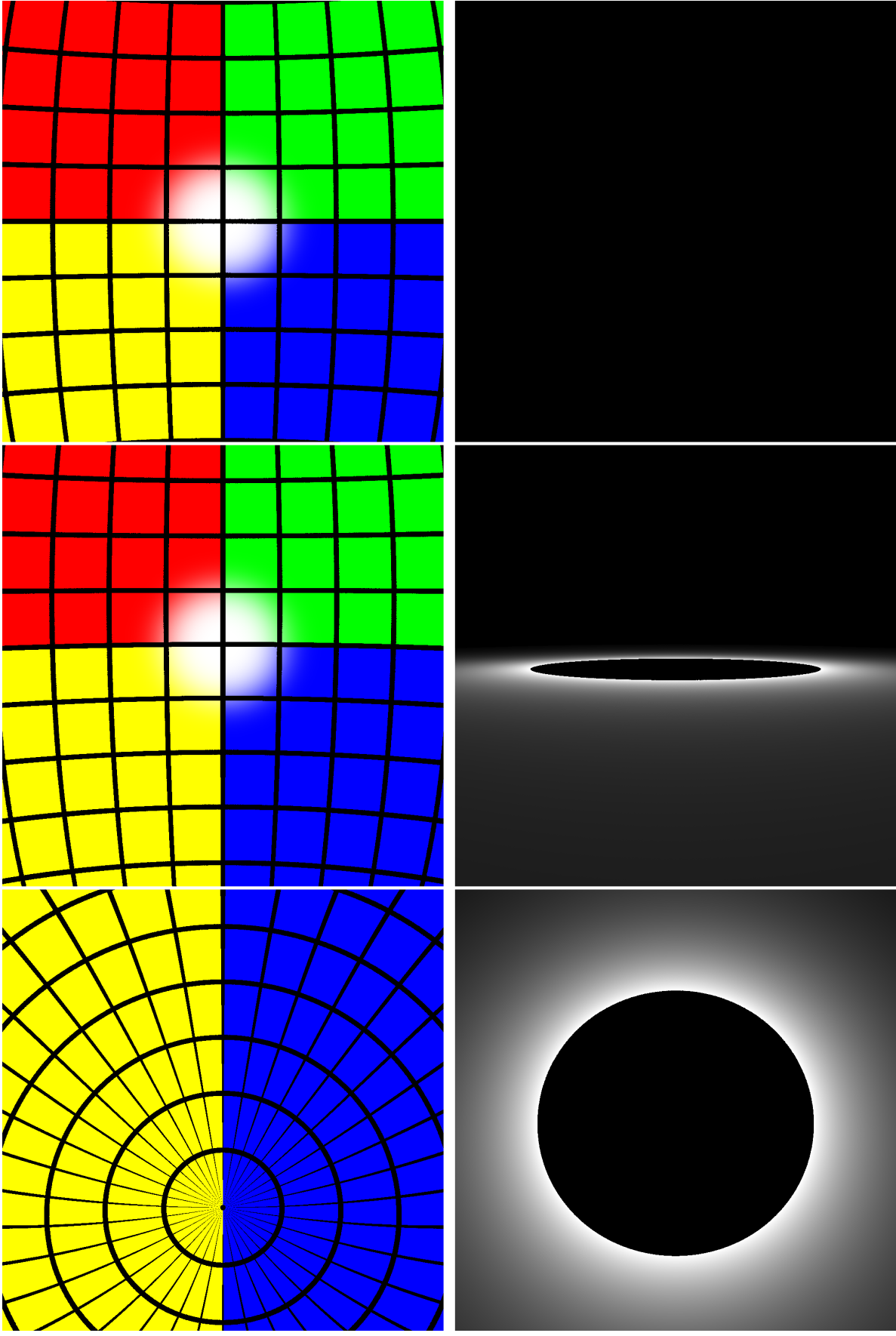}
         \caption{$w=0.98\mu$, \,\, $1^{\textrm{st}}$}
         \label{fig:w98}
     \end{subfigure}
     \hfill
     \begin{subfigure}[b]{0.28\textwidth}
         \centering
         \includegraphics[width=\textwidth]{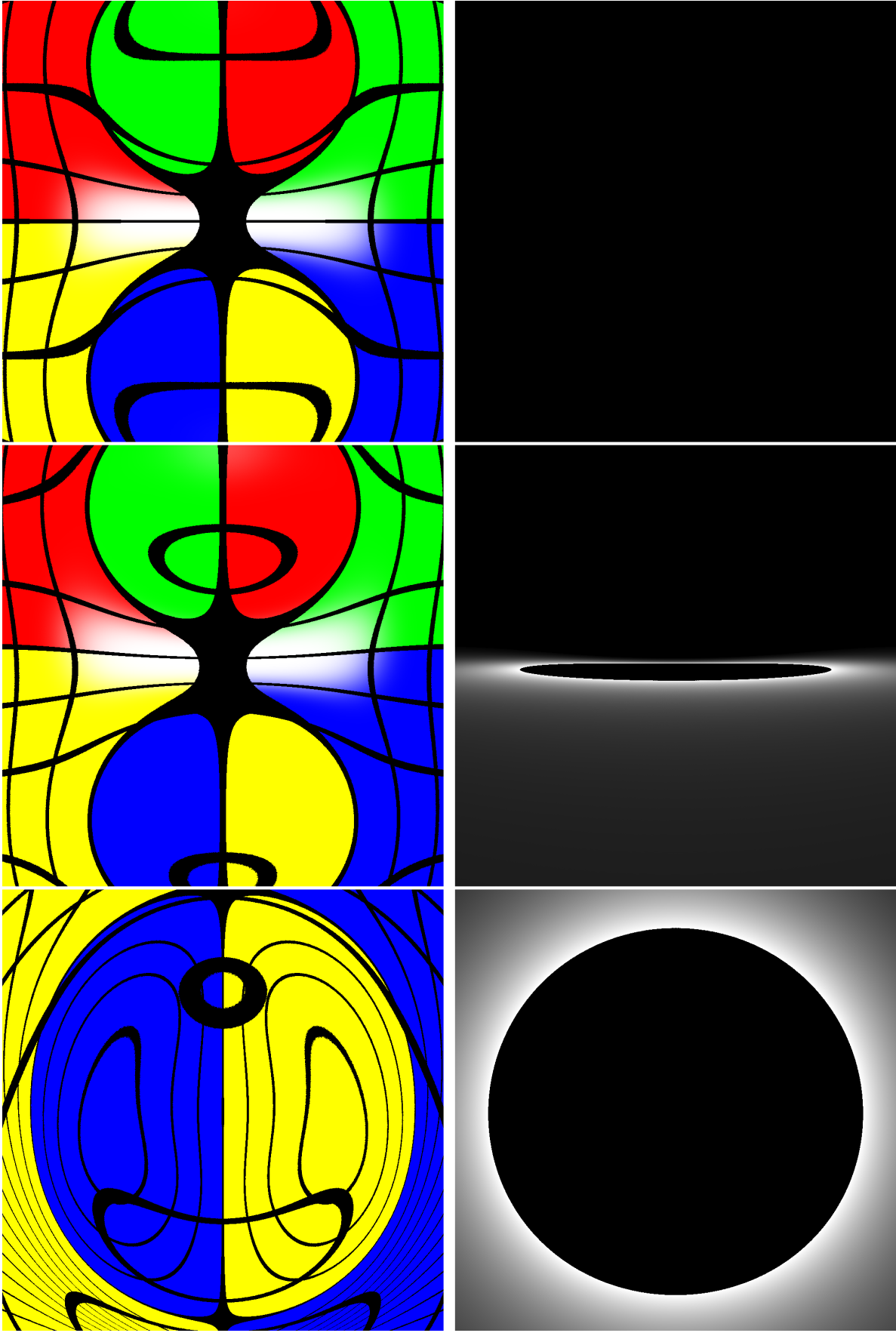}
         \caption{$\omega=0.9\mu$, \,\, $1^{\textrm{st}}$}
         \label{fig:w9}
     \end{subfigure}
     \hfill
     \begin{subfigure}[b]{0.360606061\textwidth} % multiply width of other plots by 510/396 
         \centering
         \includegraphics[width=\textwidth]{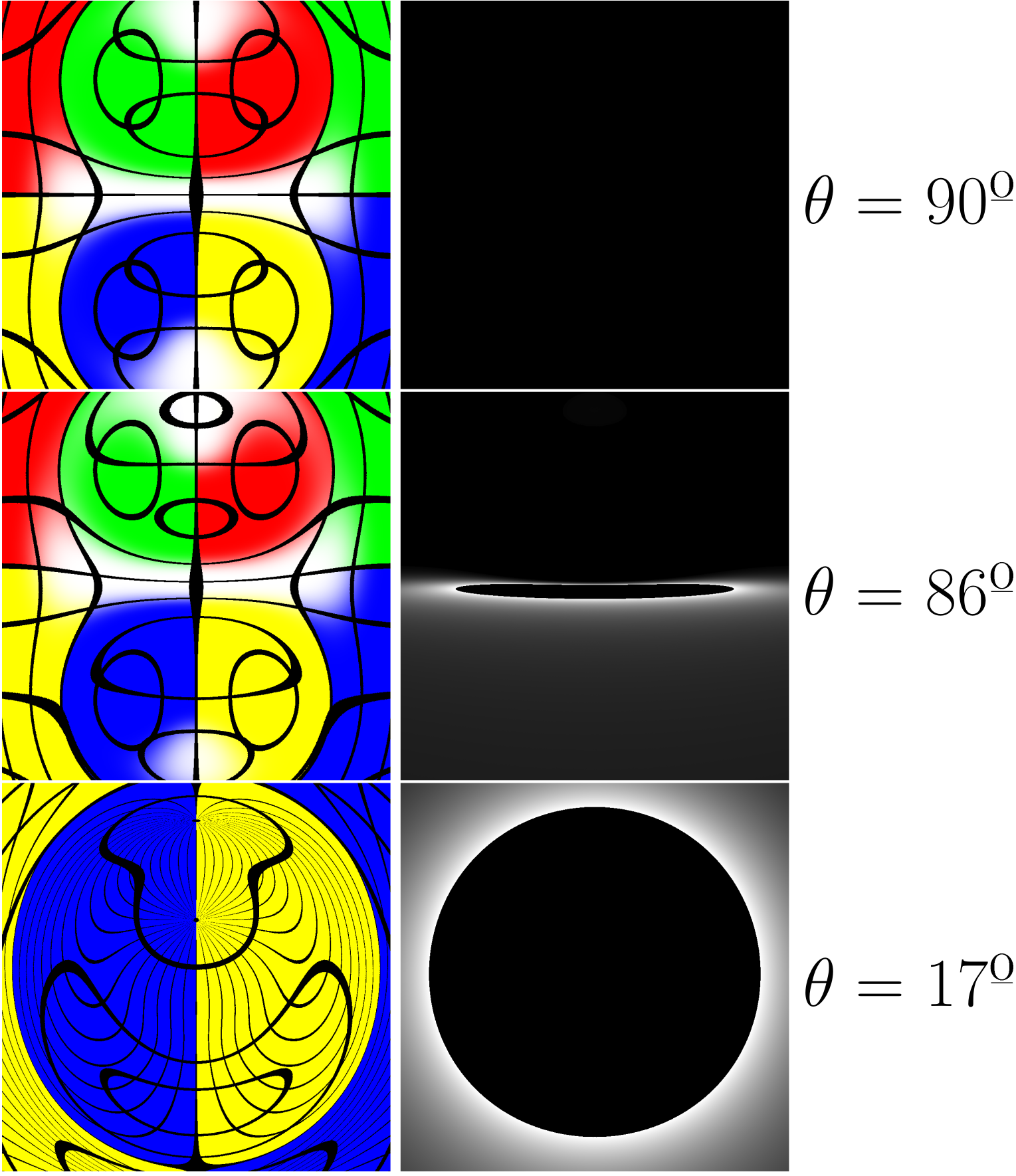}
         \caption{$\omega=0.88\mu$, \,\, $1^{\textrm{st}}$}
         \label{fig:w88}
     \end{subfigure}\\
  \vspace*{0.5cm}
     \begin{subfigure}[b]{0.28\textwidth}
         \centering
         \includegraphics[width=\textwidth]{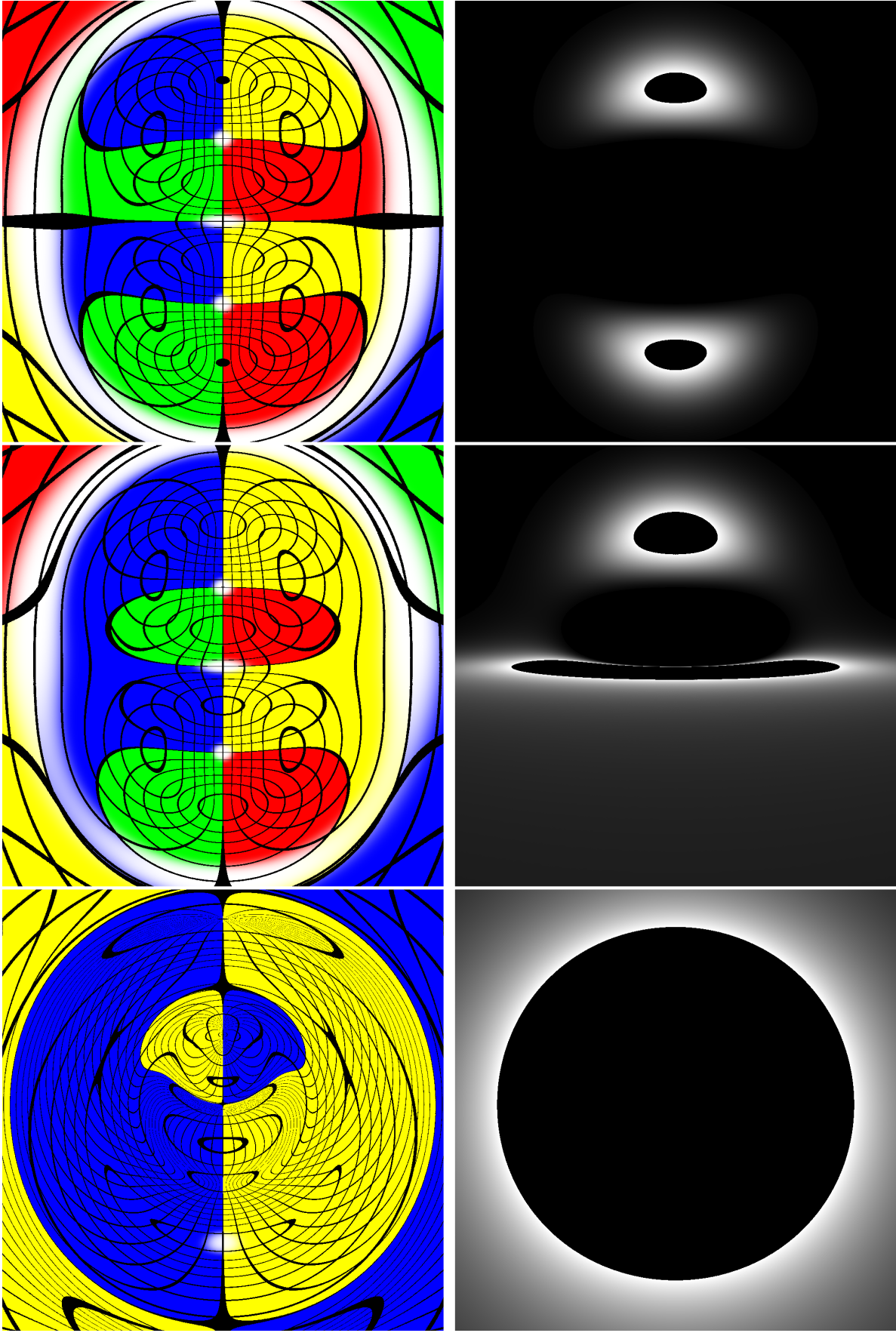}
         \caption{$\omega=0.8\mu$, \,\, $1^{\textrm{st}}$}
         \label{fig:w8}
     \end{subfigure}
     \hfill
     \begin{subfigure}[b]{0.28\textwidth}
         \centering
         \includegraphics[width=\textwidth]{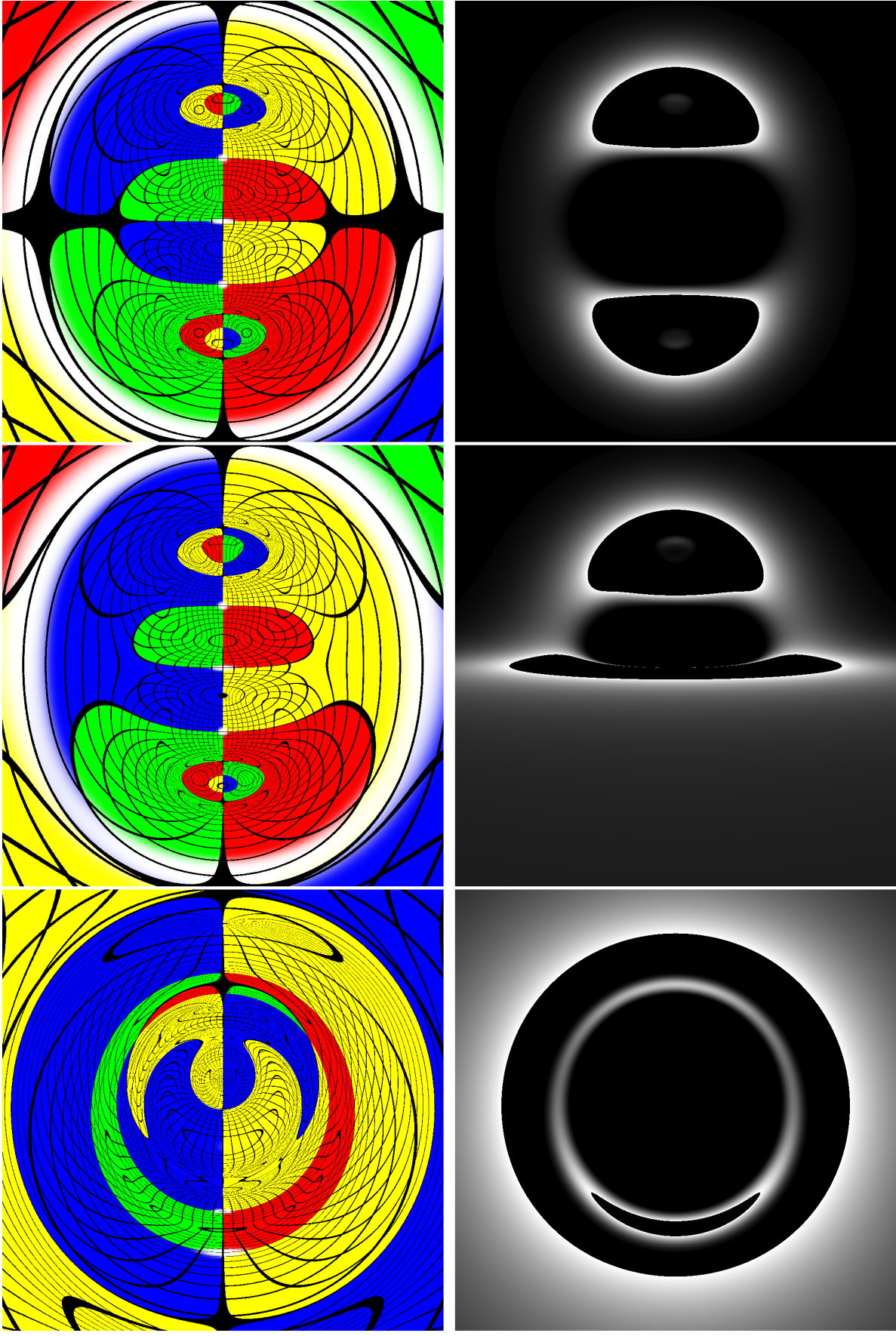}
         \caption{$\omega=0.75\mu$, \,\, $1^{\textrm{st}}$}
         \label{fig:w75}
     \end{subfigure}
     \hfill
     \begin{subfigure}[b]{0.360606061\textwidth} % multiply width of other plots by 510/396 
         \centering
         \includegraphics[width=\textwidth]{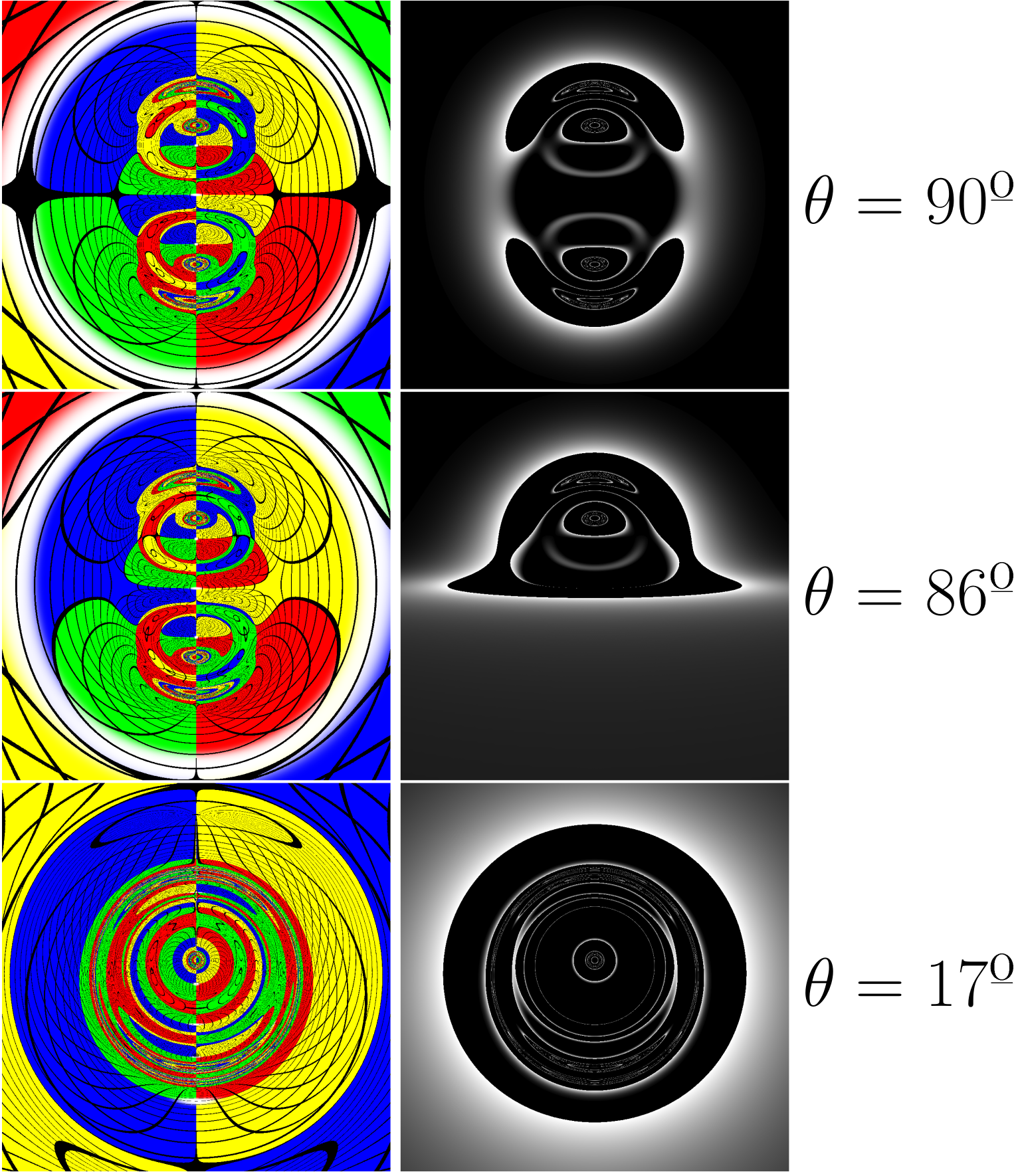}
         \caption{$\omega=0.77\mu$, \,\, $2^{\textrm{nd}}$}
         \label{fig:w77-2nd}
     \end{subfigure}
        \caption{Gravitational lensing images of DBSs displayed for different values of $\omega$ from subfigures a) to f). The left (right) column of each subfigure considers a colored sphere (equatorial disk) as the light source. Each row of every subfigure corresponds to a different observation angle $\theta$, with $\theta$ decreasing from the top to the bottom row, as indicated by the side label.}
        \label{fig:lensing}
\end{figure}

The observation images for different DBSs solutions are displayed in Figure~\ref{fig:lensing}, 
considering the two scenarios previously discussed, as well as different observation angles. The observer is set at a circumferential radius of $20M$ when in the equatorial plane. Starting from the Newtonian limit $w/\mu\sim 1$, the gravitational lensing is fairly week and fairly similar to flat spacetime (see Figure~\ref{fig:w98}). As we move along the spiral in solution space, the lensing starts to reflect the dipolar structure of the configuration. This becomes apparent in Figure~\ref{fig:w9} and Figure~\ref{fig:w88} for the colored images, but not in the accretion disk setup. However, observers close to the symmetry axis can notice an increase in the angular size of the inner edge of disk due to lensing. Further along the spiral, stronger lensing effects become more noticiable in Figure~\ref{fig:w8}, Figure~\ref{fig:w75} and Figure~\ref{fig:w77-2nd}, with dipolar structures and multiple images becoming more apparent in the disk images. Figure~\ref{fig:w77-2nd} in particular displays an intricate complex structure, characteristic of ultra-compact objects~\cite{Cunha:2016bjh}. However, no light rings were detected in the DBSs analysed (on or outside equatorial plane). But there are subtle hints there could exist Fundamental Photon Orbits~\cite{Cunha:2017eoe}, non planar generalizations of light rings, for the most compact solutions analysed. If confirmed, they would not be anchored in any Light Rings, and they would explain some of the more complex lensing features displayed. Such (novel) possibility deserves a detailed analysis that is beyond the scope of this paper. Its confirmation will yield a new application of this type of solutions, potentially paving the way towards a  general understanding on the possibility of having ultracompact objects with bound photon orbits, but without Light Rings.

%%%%%%%%%%%%%%%%%%%%%%%%%%%%%%%%%%%%%%%%%%%%%%%%%%%%%%%%%%%%%%%%

 %%%%%%%%%%%%%%%%%%%%%%%%%%%%%%%%%%%%%%%%%%%
%\section{Dynamical stability}
 %%%%%%%%%%%%%%%%%%%%%%%%%%%%%%%%%%%%%%%%%%%

 %%%%%%%%%%%%%%%%%%%%%%%%%%%%%%%%%%%%%%%%%%%
\section{Further remarks and  conclusions}
\label{sec-final}
 %%%%%%%%%%%%%%%%%%%%%%%%%%%%%%%%%%%%%%%%%%%

BSs are perhaps the simplest self-gravitating solitons. As argued in~\cite{Herdeiro:2020kvf} there is an infinite tower of multipolar solutions,  in close analogy with the hydrogen orbitals. The DBSs discussed here can be seen as two BSs in equilibrium due to a balance between gravitational attraction and scalar repulsion.

An interesting perspective on BSs is that they are a manifestation  of the
{\it 'gravitational desingularization'} mechanism  \cite{Herdeiro:2020kvf}.
Restricting to the original (and simplest) spherical solutions in 
\cite{Kaup:1968zz,Ruffini:1969qy}, one can explain this mechanism as follows.  A
scalar field on  flat spacetime (with mass $\mu$ and frequency $\omega$)
possesses a spherically symmetric solutions (of the linear KG equation) which
vanishes in the far field and diverges at $r=0$:
$$
\Phi = e^{-i \omega t} \frac{e^{-\sqrt{\mu^2 -\omega^2 } r}}{r} \ .
$$
When coupling the scalar field to gravity, 
the backreaction  regularizes this central singularity\footnote{
The same mechanism works for other models 
with more complicated matter fields \cite{Volkov:1998cc}.},
resulting in spherically symmetric BSs.
However, whereas for the flat space solution the scalar field is confined for any $\omega<\mu$,
gravity effects lead to the existence of a nonzero minimal value for the frequency. 
As argued in~\cite{Herdeiro:2020kvf},
this mechanism  works also the multipoles 
of the scalar field (see also \cite{Herdeiro:2021mol}).
The simplest case is the dipole,
with the flat spacetime solution
\begin{eqnarray}
\Phi  =
\Big(\frac{1}{r}+ \sqrt{\mu^2 -\omega^2}\Big)
 e^{-\sqrt{ \mu^2 -\omega^2 } r}
\frac{\cos\theta}{r} 
e^{-i \omega t} \ .
\end{eqnarray}
Again, gravity regularizes the singularity at $r=0$, resulting 
in physically reasonable configurations -- the DBSs - consisting of a pair of 
oscillating lumps with a phase difference of $\pi$.
It is the phase difference between these components
which
yields a repulsive
force between them, balancing the gravitational
attraction.
 
\medskip

 DBSs represent a gravitationally bound pair of solitons. Analogous configurations are known in classical field theory; the solitons possess a
core in which most energy is localized, and an asymptotic tail, which is responsible for the long-range interaction
between the well-separated solitons, see, $e.g.$ \cite{Manton:2004tk,Shnir:2018yzp}.
This interaction can be either repulsive or attractive.  
In flat space, a
pair of solitons may exist as a regular equilibrium solution
of the field equations if there are several field components and the
corresponding
interactions balance,  providing a zero net force \footnote{Notably,
there is a very special class of self-dual solitons whose  energy of interaction is always zero
for any separation between the solitons.} - see \cite{Shnir:2021gki} for examples in various spacetime dimensions.
For instance, bound pairs
of solitons may exist in the 3+1 dimensional Skyrme model with a special form of the potential,
which combines both repulsive and attractive interactions \cite{Gillard:2015eia,Gudnason:2016tiz}. Similarly,
the Faddeev-Skyrme model supports  bound pairs of Hopfions \cite{Ward:2000qj}.
Another type of bounded soliton solutions is a static pair of a Skyrmion and an anti-Skyrmion in equilibrium
\cite{Krusch:2004uf,Shnir:2009ct,Gibbons:2010cr}. Similar soliton-antisoliton pairs exist in the $SU(2)$
Yang-Mills-Higgs theory \cite{Kleihaus:1999sx,Kleihaus:2003nj,Kleihaus:2004is}.
The magnetic dipole is a  saddle point solution where the attractive short-range
forces, mediated both by the $A^3_\mu$ vector boson and the
Higgs boson, are balanced by the repulsive interaction due to massive vector bosons $A^{\pm}_\mu$
with opposite orientation in the group space \cite{Shnir:2005te}. The effective net potential of the
interaction between a monopole and an anti-monopole is attractive for large separation between the constituents
and it is repulsive on a short distance.
Similar bound pairs of solutions exist in the $SU(2)\times U(1)$ electroweak theory
\cite{Klinkhamer:1985ki,Klinkhamer:1993hb,Kleihaus:2008gn}, in the Goldstone model with
an isovector scalar field \cite{Paturyan:2005ik} and in the Euclidian $SU(2)$ Yang--Mills theory \cite{Radu:2006gg}.
These axially-symmetric  configurations represent a deformation of the topologically trivial sector,
a saddle point solution (sphaleron).
Coupling to gravity yields an additional attractive effect in the system, introducing more intricate patterns in the solution space of self-gravitating bound pairs of solitons, see
\cite{Kleihaus:2004fh,Kunz:2007jw,Ibadov:2008hj,Shnir:2015aba,Shnir:2020hau}.

\medskip

The main purpose of this paper was to provide
a deeper study of DBS solutions. 
This type of solutions   likely provides
the simplest example of composite configurations in a field theory
coupled
with gravity. In (electro-)vacuum GR, 
the two BH solution is
plagued by conical singulaties \cite{BW}, except in the extremal case, that yields a no-force condition and connects to supersymmetry and BPS bounds~\cite{Gibbons:1982fy,Tod:1983pm}.
The existence of the DBSs shows that the situation 
can be different for (rather simple) field theory models, also admitting self-gravitating solitons.

The simple effective model based on~(\ref{emodel}), managed to reproduce some features of DBSs. However, it suggests that these solutions could be dynamically stable, whereas the study in~\cite{Sanchis-Gual:2021edp} has unveiled they are unstable. This is most likely due to the point particle approximation. The two centres in the DBSs are extended objects with a more complicated dynamics. This discussion suggests, nonetheless, that the dynamical properties of DBSs can be made more robust by making each centre more compact (and hence point like). This is precisely the effect of introducing a (repulsive) self-interaction. Preliminary results indeed show this can improve the dynamical robustness of DBSs (as for other excited BSs~\cite{Siemonsen:2020hcg,Sanchis-Gual:2021phr}) and we expect to report on this soon.

If would be interesting to see if the DBSs allow for BH generalizations.
For a single, spherically symmetric BS, the answer is negative, as shown by the results in Ref. \cite{Pena:1997cy}.
Although the arguments in \cite{Pena:1997cy} do not seem to possess a simple extension for DBSs,
it is likely that also in this case there are no BHs.
For example, 
when assuming the existence of a power series expansion
of the solutions in the vicinity of the event horizon, one finds that the scalar field vanishes order by order, as implied by the regularity assumption.
However, rotation allows for a loophole,
the non-zero scalar field at the horizon being anchored to the {\it synchronization condition $\omega \sim \Omega_H$}
(with $\Omega_H$ the event horizon velocity).
Indeed, solutions of the  Einstein-scalar field equations with an event horizon inside a single spinning BS  \cite{Herdeiro:2014goa}
or a spinning DBS \cite{Kunz:2019bhm} are known to exist.

Finally, let us remark that similar   two-centre  solitonic solutions in equilibrium exist in the Einstein-(complex, massive)-Proca model~\cite{Sanchis-Gual:2021edp}, and deserve an analogous detailed study.

\section*{Acknowledgments}
This work was supported by the Center for Research
and Development in Mathematics and Applications
(CIDMA) through the Portuguese Foundation for
Science and Technology (FCT - Fundac\~ao para a
Ci\^encia e a Tecnologia), references UIDB/04106/2020
and UIDP/04106/2020, by national funds (OE),
through FCT, I.P., in the scope of the framework
contract foreseen in the numbers 4, 5 and 6 of the
article 23, of the Decree-Law 57/2016, of August
29, changed by Law 57/2017, of July 19 and by the
projects PTDC/FIS-OUT/28407/2017, CERN/FIS-
PAR/0027/2019, PTDC/FIS-AST/3041/2020 and
CERN/FIS-PAR/0024/2021. This work has further
been supported by the European Union’s Horizon 2020
research and innovation (RISE) programme H2020-
MSCA-RISE-2017 Grant No. FunFiCO-777740 and
by FCT through Project No. UIDB/00099/2020. PC
is supported by the Individual CEEC program 2020
funded by the FCT.  Computations have been performed at
the Argus and Blafis cluster at the U. Aveiro.

%%%%%%%%%%%%%%%%%%%%%%%%%%%%%%%%%%%%%%%%%%%%%%%%%%%%%%%%%%%%%%%%%%
\begin{small}
%\begin{thebibliography}{99}
\bibliographystyle{unsrt}
\bibliography{sphalerons.bib}

\begin{thebibliography}{10}

\bibitem{Weyl:1917gp}
H.~Weyl.
\newblock {The theory of gravitation}.
\newblock {\em Annalen Phys.}, 54:117--145, 1917.

\bibitem{Einstein:1936fp}
Albert Einstein and N.~Rosen.
\newblock {Two-Body Problem in General Relativity Theory}.
\newblock {\em Phys. Rev.}, 49:404--405, 1936.

\bibitem{Majumdar:1947eu}
S.~D. Majumdar.
\newblock {A class of exact solutions of Einstein's field equations}.
\newblock {\em Phys. Rev.}, 72:390--398, 1947.

\bibitem{Papapetrou:1948jw}
A.~Papapetrou.
\newblock {Einstein's theory of gravitation and flat space}.
\newblock {\em Proc. Roy. Irish Acad. A}, 52:11--23, 1948.

\bibitem{Hartle:1972ya}
J.~B. Hartle and S.~W. Hawking.
\newblock {Solutions of the Einstein-Maxwell equations with many black holes}.
\newblock {\em Commun. Math. Phys.}, 26:87--101, 1972.

\bibitem{Gibbons:1986cp}
G.~W. Gibbons and P.~J. Ruback.
\newblock {The Motion of Extreme {Reissner-Nordstrom} Black Holes in the Low
  Velocity Limit}.
\newblock {\em Phys. Rev. Lett.}, 57:1492, 1986.

\bibitem{Ferrell:1987gf}
Robert~C. Ferrell and Douglas~M. Eardley.
\newblock {Slow motion scattering and coalescence of maximally charged black
  holes}.
\newblock {\em Phys. Rev. Lett.}, 59:1617, 1987.

\bibitem{Manton:1988ba}
N.~S. Manton.
\newblock {Unstable Manifolds and Soliton Dynamics}.
\newblock {\em Phys. Rev. Lett.}, 60:1916, 1988.

\bibitem{Kaup:1968zz}
David~J. Kaup.
\newblock {Klein-Gordon Geon}.
\newblock {\em Phys. Rev.}, 172:1331--1342, 1968.

\bibitem{Ruffini:1969qy}
Remo Ruffini and Silvano Bonazzola.
\newblock {Systems of selfgravitating particles in general relativity and the
  concept of an equation of state}.
\newblock {\em Phys. Rev.}, 187:1767--1783, 1969.

\bibitem{Feinblum:1968nwc}
David~A. Feinblum and William~A. McKinley.
\newblock {Stable States of a Scalar Particle in Its Own Gravational Field}.
\newblock {\em Phys. Rev.}, 168(5):1445, 1968.

\bibitem{Herdeiro:2017fhv}
Carlos A.~R. Herdeiro, Alexandre~M. Pombo, and Eugen Radu.
\newblock {Asymptotically flat scalar, Dirac and Proca stars: discrete vs.
  continuous families of solutions}.
\newblock {\em Phys. Lett. B}, 773:654--662, 2017.

\bibitem{Sanchis-Gual:2021phr}
Nicolas Sanchis-Gual, Carlos Herdeiro, and Eugen Radu.
\newblock {Self-interactions can stabilize excited boson stars}.
\newblock {\em Class. Quant. Grav.}, 39(6):064001, 2022.

\bibitem{Herdeiro:2019mbz}
C.~Herdeiro, I.~Perapechka, E.~Radu, and Ya. Shnir.
\newblock {Asymptotically flat spinning scalar, Dirac and Proca stars}.
\newblock {\em Phys. Lett. B}, 797:134845, 2019.

\bibitem{Schunck:2003kk}
Franz~E. Schunck and Eckehard~W. Mielke.
\newblock {General relativistic boson stars}.
\newblock {\em Class. Quant. Grav.}, 20:R301--R356, 2003.

\bibitem{Liebling:2012fv}
Steven~L. Liebling and Carlos Palenzuela.
\newblock {Dynamical Boson Stars}.
\newblock {\em Living Rev.Rel.}, 15:6, 2012.

\bibitem{Shnir:2022lba}
Yakov Shnir.
\newblock {Boson Stars}.
\newblock 4 2022.

\bibitem{Yoshida:1997nd}
Shijun Yoshida and Yoshiharu Eriguchi.
\newblock {New static axisymmetric and nonvacuum solutions in general
  relativity: Equilibrium solutions of boson stars}.
\newblock {\em Phys. Rev. D}, 55:1994--2001, 1997.

\bibitem{Schupp:1995dy}
B.~Schupp and J.~J. van~der Bij.
\newblock {An axially symmetric Newtonian boson star}.
\newblock {\em Phys. Lett. B}, 366:85--88, 1996.

\bibitem{Yoshida:1997jq}
Shijun Yoshida and Yoshiharu Eriguchi.
\newblock {Nonaxisymmetric boson stars in Newtonian gravity}.
\newblock {\em Phys. Rev. D}, 56:6370--6377, 1997.

\bibitem{Herdeiro:2020kvf}
C.~A.~R. Herdeiro, J.~Kunz, I.~Perapechka, E.~Radu, and Ya. Shnir.
\newblock {Multipolar boson stars: macroscopic Bose-Einstein condensates akin
  to hydrogen orbitals}.
\newblock {\em Phys. Lett. B}, 812:136027, 2021.

\bibitem{Herdeiro:2021mol}
C.~A.~R. Herdeiro, J.~Kunz, I.~Perapechka, E.~Radu, and Y.~Shnir.
\newblock {Chains of Boson Stars}.
\newblock {\em Phys. Rev. D}, 103(6):065009, 2021.

\bibitem{Gervalle:2022fze}
Romain Gervalle.
\newblock {Chains of rotating boson stars}.
\newblock {\em Phys. Rev. D}, 105:124052, 2022.

\bibitem{Palenzuela:2006wp}
C.~Palenzuela, I.~Olabarrieta, L.~Lehner, and Steven~L. Liebling.
\newblock {Head-on collisions of boson stars}.
\newblock {\em Phys. Rev. D}, 75:064005, 2007.

\bibitem{Stephani:2003tm}
Hans Stephani, D.~Kramer, Malcolm A.~H. MacCallum, Cornelius Hoenselaers, and
  Eduard Herlt.
\newblock {\em {Exact solutions of Einstein's field equations}}.
\newblock Cambridge Monographs on Mathematical Physics. Cambridge Univ. Press,
  Cambridge, 2003.

\bibitem{Herdeiro:2015gia}
Carlos Herdeiro and Eugen Radu.
\newblock {Construction and physical properties of Kerr black holes with scalar
  hair}.
\newblock {\em Class. Quant. Grav.}, 32(14):144001, 2015.

\bibitem{Wald:1984rg}
Robert~M. Wald.
\newblock {\em {General Relativity}}.
\newblock Chicago Univ. Pr., Chicago, USA, 1984.

\bibitem{Amaro-Seoane:2010pks}
Pau Amaro-Seoane, Juan Barranco, Argelia Bernal, and Luciano Rezzolla.
\newblock {Constraining scalar fields with stellar kinematics and collisional
  dark matter}.
\newblock {\em JCAP}, 11:002, 2010.

\bibitem{Wiseman:2002zc}
Toby Wiseman.
\newblock {Static axisymmetric vacuum solutions and nonuniform black strings}.
\newblock {\em Class. Quant. Grav.}, 20:1137--1176, 2003.

\bibitem{schoen}
W~Sch{\"o}nauer and R~Wei$\beta$.
\newblock Efficient vectorizable pde solvers.
\newblock {\em Journal of computational and applied mathematics},
  27(1-2):279--297, 1989.

\bibitem{schauder1992cadsol}
M~Schauder, R~Wei{\ss}, and W~Sch{\"o}nauer.
\newblock The cadsol program package.
\newblock {\em Universit{\"a}t Karlsruhe, Interner Bericht}, (46/92):26, 1992.

\bibitem{pardiso}
Nicholas~IM Gould, Jennifer~A Scott, and Yifan Hu.
\newblock A numerical evaluation of sparse direct solvers for the solution of
  large sparse symmetric linear systems of equations.
\newblock {\em ACM Transactions on Mathematical Software (TOMS)}, 33(2):10--es,
  2007.

\bibitem{schenk}
Olaf Schenk and Klaus G{\"a}rtner.
\newblock Solving unsymmetric sparse systems of linear equations with pardiso.
\newblock {\em Future Generation Computer Systems}, 20(3):475--487, 2004.

\bibitem{Friedberg:1986tp}
R.~Friedberg, T.~D. Lee, and Y.~Pang.
\newblock {MINI - SOLITON STARS}.
\newblock {\em Phys. Rev. D}, 35:3640, 1987.

\bibitem{Coleman:1985ki}
Sidney~R. Coleman.
\newblock {Q-balls}.
\newblock {\em Nucl. Phys. B}, 262(2):263, 1985.
\newblock [Addendum: Nucl.Phys.B 269, 744 (1986)].

\bibitem{Gibbons:2010cr}
G.~W. Gibbons, C.~M. Warnick, and W.~W. Wong.
\newblock {Non-existence of Skyrmion-Skyrmion and Skrymion-anti-Skyrmion static
  equilibria}.
\newblock {\em J. Math. Phys.}, 52:012905, 2011.

\bibitem{Volkov:2002aj}
Mikhail~S. Volkov and Erik Wohnert.
\newblock {Spinning Q balls}.
\newblock {\em Phys. Rev. D}, 66:085003, 2002.

\bibitem{Kleihaus:2007vk}
Burkhard Kleihaus, Jutta Kunz, Meike List, and Isabell Schaffer.
\newblock {Rotating Boson Stars and Q-Balls. II. Negative Parity and
  Ergoregions}.
\newblock {\em Phys. Rev. D}, 77:064025, 2008.

\bibitem{Cunha:2018acu}
Pedro V.~P. Cunha and Carlos A.~R. Herdeiro.
\newblock {Shadows and strong gravitational lensing: a brief review}.
\newblock {\em Gen. Rel. Grav.}, 50(4):42, 2018.

\bibitem{Bohn:2014xxa}
Andy Bohn, William Throwe, Fran H\'ebert, Katherine Henriksson, Darius
  Bunandar, Mark~A. Scheel, and Nicholas~W. Taylor.
\newblock {What does a binary black hole merger look like?}
\newblock {\em Class. Quant. Grav.}, 32(6):065002, 2015.

\bibitem{Cunha:2016bpi}
Pedro V.~P. Cunha, Carlos A.~R. Herdeiro, Eugen Radu, and Helgi~F. Runarsson.
\newblock {Shadows of Kerr black holes with and without scalar hair}.
\newblock {\em Int. J. Mod. Phys. D}, 25(09):1641021, 2016.

\bibitem{Cunha:2015yba}
Pedro V.~P. Cunha, Carlos A.~R. Herdeiro, Eugen Radu, and Helgi~F. Runarsson.
\newblock {Shadows of Kerr black holes with scalar hair}.
\newblock {\em Phys. Rev. Lett.}, 115(21):211102, 2015.

\bibitem{Cunha:2018gql}
Pedro V.~P. Cunha, Carlos A.~R. Herdeiro, and Maria~J. Rodriguez.
\newblock {Does the black hole shadow probe the event horizon geometry?}
\newblock {\em Phys. Rev. D}, 97(8):084020, 2018.

\bibitem{Cunha:2018cof}
Pedro V.~P. Cunha, Carlos A.~R. Herdeiro, and Maria~J. Rodriguez.
\newblock {Shadows of Exact Binary Black Holes}.
\newblock {\em Phys. Rev. D}, 98(4):044053, 2018.

\bibitem{Cunha:2019hzj}
Pedro V.~P. Cunha, Nelson~A. Eir\'o, Carlos A.~R. Herdeiro, and Jos\'e P.~S.
  Lemos.
\newblock {Lensing and shadow of a black hole surrounded by a heavy accretion
  disk}.
\newblock {\em JCAP}, 03:035, 2020.

\bibitem{Herdeiro:2021lwl}
Carlos A.~R. Herdeiro, Alexandre~M. Pombo, Eugen Radu, Pedro V.~P. Cunha, and
  Nicolas Sanchis-Gual.
\newblock {The imitation game: Proca stars that can mimic the Schwarzschild
  shadow}.
\newblock {\em JCAP}, 04:051, 2021.

\bibitem{Cunha:2016bjh}
P.~V.~P. Cunha, J.~Grover, C.~Herdeiro, E.~Radu, H.~Runarsson, and A.~Wittig.
\newblock {Chaotic lensing around boson stars and Kerr black holes with scalar
  hair}.
\newblock {\em Phys. Rev. D}, 94(10):104023, 2016.

\bibitem{Cunha:2017eoe}
Pedro V.~P. Cunha, Carlos A.~R. Herdeiro, and Eugen Radu.
\newblock {Fundamental photon orbits: black hole shadows and spacetime
  instabilities}.
\newblock {\em Phys. Rev. D}, 96(2):024039, 2017.

\bibitem{Volkov:1998cc}
Mikhail~S. Volkov and Dmitri~V. Gal'tsov.
\newblock {Gravitating nonAbelian solitons and black holes with Yang-Mills
  fields}.
\newblock {\em Phys. Rept.}, 319:1--83, 1999.

\bibitem{Manton:2004tk}
N.~S. Manton and P.~Sutcliffe.
\newblock {\em {Topological solitons}}.
\newblock Cambridge Monographs on Mathematical Physics. Cambridge University
  Press, 2004.

\bibitem{Shnir:2018yzp}
Yakov~M. Shnir.
\newblock {\em {Topological and Non-Topological Solitons in Scalar Field
  Theories}}.
\newblock Cambridge University Press, 7 2018.

\bibitem{Shnir:2021gki}
Yakov~M. Shnir.
\newblock {Chains of Interacting Solitons}.
\newblock {\em Symmetry}, 13(2):284, 2021.

\bibitem{Gillard:2015eia}
Mike Gillard, Derek Harland, and Martin Speight.
\newblock {Skyrmions with low binding energies}.
\newblock {\em Nucl. Phys. B}, 895:272--287, 2015.

\bibitem{Gudnason:2016tiz}
Sven~Bjarke Gudnason, Baiyang Zhang, and Nana Ma.
\newblock {Generalized Skyrme model with the loosely bound potential}.
\newblock {\em Phys. Rev. D}, 94(12):125004, 2016.

\bibitem{Ward:2000qj}
R.~S. Ward.
\newblock {The Interaction of two Hopf solitons}.
\newblock {\em Phys. Lett. B}, 473:291--296, 2000.

\bibitem{Krusch:2004uf}
Steffen Krusch and Paul Sutcliffe.
\newblock {Sphalerons in the Skyrme model}.
\newblock {\em J. Phys. A}, 37:9037, 2004.

\bibitem{Shnir:2009ct}
Ya. Shnir and D.~H. Tchrakian.
\newblock {Skyrmion-Anti-Skyrmion Chains}.
\newblock {\em J. Phys. A}, 43:025401, 2010.

\bibitem{Kleihaus:1999sx}
Burkhard Kleihaus and Jutta Kunz.
\newblock {A Monopole - anti-monopole solution of the SU(2) Yang-Mills-Higgs
  model}.
\newblock {\em Phys. Rev. D}, 61:025003, 2000.

\bibitem{Kleihaus:2003nj}
Burkhard Kleihaus, Jutta Kunz, and Yasha Shnir.
\newblock {Monopole anti-monopole chains}.
\newblock {\em Phys. Lett. B}, 570:237--243, 2003.

\bibitem{Kleihaus:2004is}
Burkhard Kleihaus, Jutta Kunz, and Yasha Shnir.
\newblock {Monopole-antimonopole chains and vortex rings}.
\newblock {\em Phys. Rev. D}, 70:065010, 2004.

\bibitem{Shnir:2005te}
Yasha Shnir.
\newblock {Electromagnetic interaction in the system of multimonopoles and
  vortex rings}.
\newblock {\em Phys. Rev. D}, 72:055016, 2005.

\bibitem{Klinkhamer:1985ki}
Frans~R. Klinkhamer.
\newblock {A New Sphaleron in the {Weinberg-Salam} Theory?}
\newblock {\em Z. Phys. C}, 29:153, 1985.

\bibitem{Klinkhamer:1993hb}
Frans~R. Klinkhamer.
\newblock {Construction of a new electroweak sphaleron}.
\newblock {\em Nucl. Phys. B}, 410:343--354, 1993.

\bibitem{Kleihaus:2008gn}
Burkhard Kleihaus, Jutta Kunz, and Michael Leissner.
\newblock {Sphalerons, Antisphalerons and Vortex Rings}.
\newblock {\em Phys. Lett. B}, 663:438--444, 2008.

\bibitem{Paturyan:2005ik}
Vanush Paturyan, Eugen Radu, and D.~H. Tchrakian.
\newblock {Solitons and soliton-antisoliton pairs of a goldstone model in 3+1
  dimensions}.
\newblock {\em J. Phys. A}, 39:3817--3828, 2006.

\bibitem{Radu:2006gg}
Eugen Radu and D.~H. Tchrakian.
\newblock {Self-dual instanton and nonself-dual instanton-antiinstanton
  solutions in d=4 Yang-Mills theory}.
\newblock {\em Phys. Lett. B}, 636:201--206, 2006.

\bibitem{Kleihaus:2004fh}
Burkhard Kleihaus, Jutta Kunz, and Yasha Shnir.
\newblock {Gravitating monopole-antimonopole chains and vortex rings}.
\newblock {\em Phys. Rev. D}, 71:024013, 2005.

\bibitem{Kunz:2007jw}
Jutta Kunz, Ulrike Neemann, and Yasha Shnir.
\newblock {Gravitating monopole-antimonopole systems at large scalar coupling}.
\newblock {\em Phys. Rev. D}, 75:125008, 2007.

\bibitem{Ibadov:2008hj}
Rustam Ibadov, Burkhard Kleihaus, Jutta Kunz, and Michael Leissner.
\newblock {Gravitating Sphaleron-Antisphaleron Systems}.
\newblock {\em Phys. Lett. B}, 663:136--140, 2008.

\bibitem{Shnir:2015aba}
Ya. Shnir.
\newblock {Gravitating sphalerons in the Skyrme model}.
\newblock {\em Phys. Rev. D}, 92(8):085039, 2015.

\bibitem{Shnir:2020hau}
Yakov Shnir.
\newblock {Black holes with Skyrmion-anti-Skyrmion hairs}.
\newblock {\em Phys. Lett. B}, 810:135847, 2020.

\bibitem{BW}
Rudolf Bach and Hermann Weyl.
\newblock {Neue Lösungen der Einsteinschen Gravitationsgleichungen. B.
  Explicite Aufstellung statischer axialsymmetrischer Felder. Mit einem Zusatz
  über das statische Zweikörperproblem von H. Weyl}.
\newblock {\em Mathematische Zeitschrift}, 13:134--145, 1922.

\bibitem{Gibbons:1982fy}
G.~W. Gibbons and C.~M. Hull.
\newblock {A Bogomolny Bound for General Relativity and Solitons in N=2
  Supergravity}.
\newblock {\em Phys. Lett. B}, 109:190--194, 1982.

\bibitem{Tod:1983pm}
K.~P. Tod.
\newblock {All Metrics Admitting Supercovariantly Constant Spinors}.
\newblock {\em Phys. Lett. B}, 121:241--244, 1983.

\bibitem{Sanchis-Gual:2021edp}
Nicolas Sanchis-Gual, Fabrizio Di~Giovanni, Carlos Herdeiro, Eugen Radu, and
  Jos\'e~A. Font.
\newblock {Multifield, Multifrequency Bosonic Stars and a Stabilization
  Mechanism}.
\newblock {\em Phys. Rev. Lett.}, 126(24):241105, 2021.

\bibitem{Siemonsen:2020hcg}
Nils Siemonsen and William~E. East.
\newblock {Stability of rotating scalar boson stars with nonlinear
  interactions}.
\newblock {\em Phys. Rev. D}, 103(4):044022, 2021.

\bibitem{Pena:1997cy}
I.~Pena and D.~Sudarsky.
\newblock {Do collapsed boson stars result in new types of black holes?}
\newblock {\em Class. Quant. Grav.}, 14:3131--3134, 1997.

\bibitem{Herdeiro:2014goa}
Carlos A.~R. Herdeiro and Eugen Radu.
\newblock {Kerr black holes with scalar hair}.
\newblock {\em Phys. Rev. Lett.}, 112:221101, 2014.

\bibitem{Kunz:2019bhm}
J.~Kunz, I.~Perapechka, and Ya. Shnir.
\newblock {Kerr black holes with parity-odd scalar hair}.
\newblock {\em Phys. Rev. D}, 100(6):064032, 2019.

\end{thebibliography}
%\printbibliography
%\bibliography{sphalerons.bib}
%%%%%%%%%%%%%%%%%%%%%%%%%%%%%%%%%%%%%%%%%%%%%%%%%%%%%%%%%%%%%%%%%% 
\end{small}

 \end{document}